\documentclass[%
reprint,
superscriptaddress,
 amsmath,amssymb,
 aps,
]{revtex4-2}

\usepackage[utf8]{inputenc}
\usepackage{siunitx}
\usepackage{array}
\usepackage{physics}
\usepackage{dsfont}
\usepackage[english]{babel}			
\usepackage{xcolor}
\usepackage{graphicx}				
\usepackage{amssymb,amsmath}		
\usepackage{url}					
\usepackage{marvosym}				
\usepackage[T1]{fontenc}            %
\usepackage{wrapfig}				
\usepackage{charter} 				
\usepackage{blindtext}				
\usepackage{datetime}				
\usepackage{lipsum}                 
\usepackage{float}                  
\usepackage{tabularx}
\usepackage{dcolumn}
\usepackage{bm}
\usepackage{todonotes}
\usepackage{hyperref}
\hypersetup{colorlinks=true,allcolors=blue}
\newcommand{\q}{\mathbf{q}}

\newcommand{\UIBK}{Institut f{\"u}r Experimentalphysik, Universit{\"a}t Innsbruck, Innsbruck, Austria}
\newcommand{\WWU}{Institut f{\"u}r Festk{\"o}rpertheorie, Universit{\"a}t M{\"u}nster, 48149 M{\"u}nster, Germany}
\newcommand{\TUdo}{Condensed Matter Theory, Department of Physics, TU Dortmund, 44221 Dortmund, Germany}
\newcommand{\Bayreuth}{Theoretische Physik III, Universit{\"a}t Bayreuth, 95440 Bayreuth, Germany}
\newcommand{\JKU}{Institute of Semiconductor and Solid State Physics, Johannes Kepler University Linz, Linz, Austria}

\preprint{APS/123-QED}

\begin{document}

\title{Collective Excitation of Spatio-Spectrally Distinct Quantum Dots Enabled by Chirped Pulses}
\author{Florian Kappe}
    \thanks{These two authors contributed equally}
    \affiliation{\UIBK}
\author{Yusuf Karli}
    \thanks{These two authors contributed equally}
    \affiliation{\UIBK}
\author{Thomas K. Bracht}%
\affiliation{\WWU}
\affiliation{\TUdo}
\author{Saimon Covre da Silva}%
\affiliation{\JKU}
\author{Tim Seidelmann}%
\affiliation{\Bayreuth}
\author{Vollrath Martin Axt}%
\affiliation{\Bayreuth}
\author{Armando Rastelli}%
\affiliation{\JKU}
\author{Gregor Weihs}%
\affiliation{\UIBK}
\author{Doris E. Reiter}%
\affiliation{\TUdo}
\author{Vikas Remesh}%
\email{vikas.remesh@uibk.ac.at}

\affiliation{\UIBK}

\date{\today}

\begin{abstract}
For a scalable photonic device producing entangled photons, it is desirable to have multiple quantum emitters in an ensemble that can be collectively excited, despite their spectral variability. For quantum dots, Rabi rotation, the most popular method for resonant excitation, cannot assure a universal, highly efficient excited state preparation, because of its sensitivity to the excitation parameters. In contrast, Adiabatic Rapid Passage (ARP), relying on chirped optical pulses, is immune to quantum dot spectral inhomogeneity. Here, we advocate the robustness of ARP for simultaneous excitation of the biexciton states of multiple quantum dots. For positive chirps, we find that there is also regime of phonon advantage that widens the tolerance range of spectral detunings. Using the same laser pulse we demonstrate the simultaneous excitation of energetically and spatially distinct quantum dots. Being able to generate spatially multiplexed entangled photon pairs is a big step towards the scalability of photonic devices.

\end{abstract}
           
\maketitle

\section*{Introduction}
Entangled photon pairs that are encoded in polarization \cite{benson_regulated_2000,akopian_entangled_2006,muller_-demand_2014,basso_basset_entanglement_2019,schimpf2021entanglement} or time-bins \cite{jayakumar_time-bin_2014,lee_quantum_2019}, are key to testing the foundations of quantum mechanics \cite{weihs_violation_1998} and implementing secure quantum key distribution protocols \cite{jennewein_quantum_2000}. 
To construct a scalable quantum photonic architecture for such purposes, it is desirable to have an ensemble of quantum emitters that can be collectively manipulated \cite{uppu_quantum-dot-based_2021,zhai_quantum_2022}. 
Semiconductor quantum dots are high brightness sources for photon generation \cite{tomm_bright_2021,liu_solid-state_2019,vajner_quantum_2022,senellart_high-performance_2017}, accomplishing high degrees of entanglement \cite{basso_basset_entanglement_2019,schimpf_quantum_2021,jayakumar_time-bin_2014,huber_coherence_2016}, low multiphoton rate \cite{hanschke_quantum_2018} and near-deterministic operating nature \cite{ding_-demand_2016}. Despite these achievements, the fabrication of quantum dots via self-assembly growth processes inevitably introduces some variability in physical properties like size, alloy and strain distribution, as well as distribution of impurities in the environment, resulting in inhomogeneous broadening and spectral diffusion of emission lines.
A possible solution could be electric, magnetic or strain field tuning \cite{zhao_advanced_2020} to manipulate the emission properties, but tuning on length scales comparable to the inter-dot distance is still a colossal task \cite{grim_scalable_2019}.
Hence, an emerging question is how to develop an optical excitation scheme that excites spatially and energetically distinct quantum dot sources of entangled photon pairs.

The standard approach of resonant two-photon excitation (TPE) of the biexciton state \cite{ramsay_review_2010,stufler_two-photon_2006,jayakumar_deterministic_2013} is disadvantageous, because of its sensitivity to the transition dipole moment of individual quantum dots, and fluctuations in laser parameters like intensity and frequency. Accordingly, in reality, a tailored laser pulse that fully inverts the population in one quantum dot will likely not do so in another. 

This calls for a robust excitation scheme, i.e., one whose efficiency is insensitive against small fluctuations in the excitation parameters. One such scheme is based on chirped excitation via Adiabatic Rapid Passage (ARP), which already has been demonstrated on single quantum dots \cite{simon_robust_2011,wu_population_2011,debnath_high-fidelity_2013,kaldewey_coherent_2017,mathew_subpicosecond_2014}.
For the production of single photons, ARP can work simultaneously on several quantum dots \cite{creatore_creation_2012,gamouras2013simultaneous,ramachandran_experimental_2021}, where just the exciton was addressed. 

In this paper, we demonstrate the simultaneous preparation of the biexciton states in a quantum dot ensemble via ARP without modifying the excitation parameters. 
Interestingly, we also find a regime of \textit{phonon advantage}, where a combination of positively chirped excitation and phonon-assisted state preparation provides an extended spectral range for high efficiency excitation of biexciton states in multiple quantum dots at moderate pulse areas. 
This is different from a pure phonon assisted excitation \cite{reiter_phonon_2012,ardelt_dissipative_2014,bounouar_phonon-assisted_2015,quilter_phonon-assisted_2015,reindl_phonon-assisted_2017,glassl_proposed_2013,thomas_bright_2021}, as it requires high pulse areas. Our results pave the way for the realization of spatially multiplexed entangled photon pairs from the same source. 

\section*{Pulse chirping} \label{chirptheory}
\begin{figure*}[!t]
    \centering
    \includegraphics[width=\linewidth]{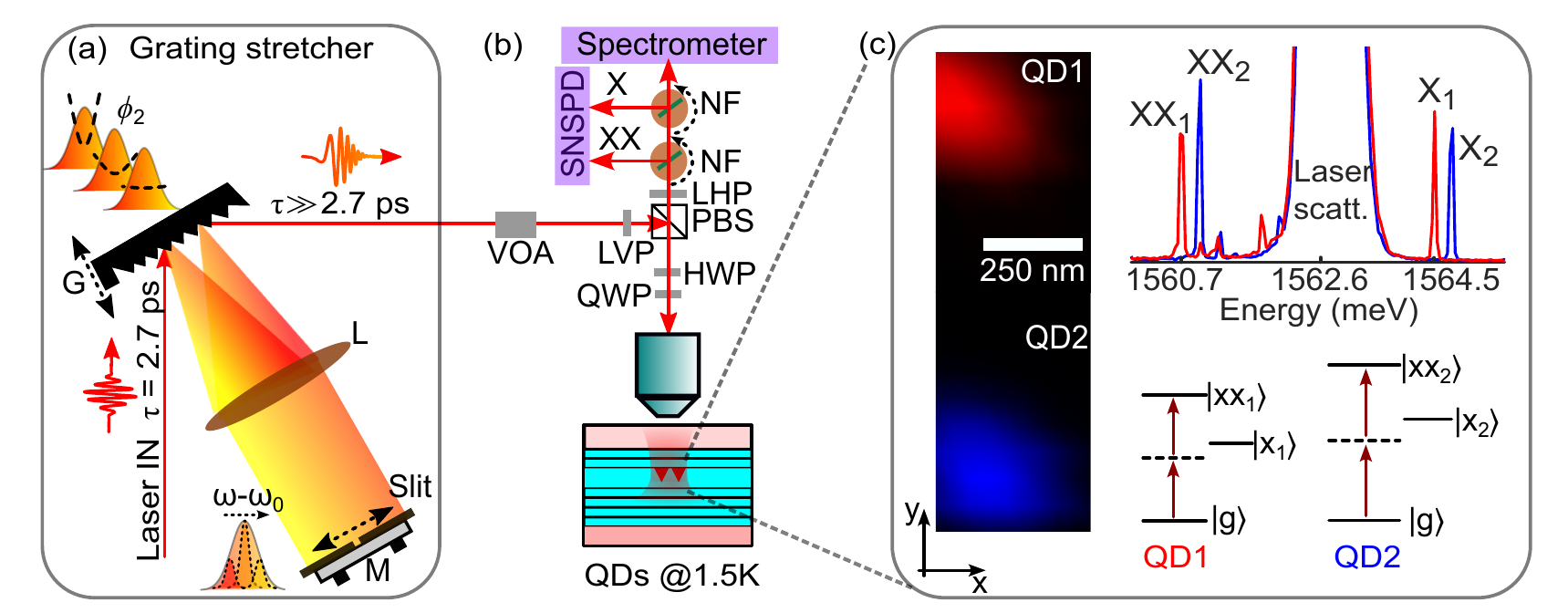}
    \caption{\textbf{Experiment overview}: (a) \textbf{Dispersion control}: The folded grating stretcher consisting of a ruled grating (G), a lens (L) and a folding mirror (M). A motorized slit mounted on the folding mirror enables laser spectral tuning. The GDD values are varied by translating the grating from the focal plane of the lens. (b) \textbf{Cryo-microscopy setup}: The excitation laser pulses are sent through a variable optical attenuator (VOA) to control pulse power and through the polarizing optics (LVP- Linear Vertical Polarizer, PBS- Polarizing Beamsplitter, HWP- Half Wave Plate, QWP- Quarter Wave Plate) towards the optical cryostat that holds the quantum dot ensemble at \SI{1.5}{\kelvin}. The emitted photons are collected via the same path, cross-polarization filtered with Linear Horizontal Polarizer (LHP) and sent either towards the spectrometer or through a home-built monochromator equipped with notch filters (NF) to the superconducting nanowire single photon detection (SNSPD) channels to record the exciton (X) and biexciton (XX) photon coincidences. (c) \textbf{Characterization}: The two chosen quantum dots are labelled QD1 and QD2, with their raster scan TPE images, colour-coded red (QD1) and blue (QD2) respectively, alongside their energy level diagrams. Also displayed are their resonant TPE emission spectra, showing spectrally shifted X and XX lines from the scattered resonant laser light.}
    \label{fig:setup}
\end{figure*}

To achieve ARP excitation, we require laser pulses of time-varying frequency (chirping). In the frequency domain this is described by
\begin{equation*}
    E(\omega)=E_{0} \exp \left[-\frac{\left(\omega-\omega_{c}\right)^{2}}{\Delta \omega^{2}}\right]\exp \left[i \frac{\phi_{2}}{2}\left(\omega-\omega_{c}\right)^{2}\right],
\end{equation*}
where $E_{0}$ is the amplitude of the Gaussian frequency envelope centered at $\omega_{c}$ with a frequency bandwidth of $\Delta \omega$ and $\phi_{2}$ denotes the group delay dispersion (GDD) or in general, the linear chirp. 
Introducing $\phi_{2}$ has two effects: it stretches the temporal duration of the laser pulse from the transform limit ($\tau_{0}$) to $\tau_{p}$ according to the relation $\tau_{p}^{2}=\tau_{0}^{2}+\left(\frac{4 \ln2\left(\phi_{2}\right)}{\tau_{0}}\right)^{2}$, and dictates the frequency ordering in the pulse. For positive $\phi_{2}$, the frequency increases over time, meaning the red part of the spectrum arrives before the blue part and vice versa for negative $\phi_{2}$.

The experimental implementation of the chirping process is sketched in Fig.~\ref{fig:setup}(a). Initially, a Ti:Sapphire laser producing pulses of \SI{2.6}{ps} time duration (measured as intensity FWHM, Tsunami, SpectraPhysics), is tuned to a wavelength of \SI{793}{nm}. 
To render $\phi_{2}$ we rely on a folded grating stretcher, that consists of a diffraction grating (1200 lines/mm, Newport) to disperse the beam, a lens ($f= \SI{750}{\milli\meter}$) to focus the spectral components to its Fourier plane where a folding mirror is mounted  \cite{backus_high_1998,martinez_3000_1987,lai_single-grating_1994,martinez1986grating}.
A motorized slit mounted on the folding mirror enables laser frequency tuning with respect to the TPE resonance. 
If the distance between the grating and the lens is $f$, laser pulses leave the stretcher dispersion-free, whereas the displacement of the grating towards the lens induces positive chirp \cite{martinez_3000_1987,backus_high_1998}. 
Pulse durations corresponding to various positions of the grating are characterized via spectral and nonlinear autocorrelation measurements (PulseCheck, APE GmbH).  
The maximum average laser power measured before the entrance of the stretcher is \SI{400}{\milli\watt}, which corresponds to $\approx$ \SI{5}{\nano\joule} energy per pulse and a peak pulse power of $\approx$ \SI{1.8}{\kilo\watt}. 

\section*{ARP Process}

\begin{figure}[!hbt]
    \centering
    \includegraphics[width =\linewidth]{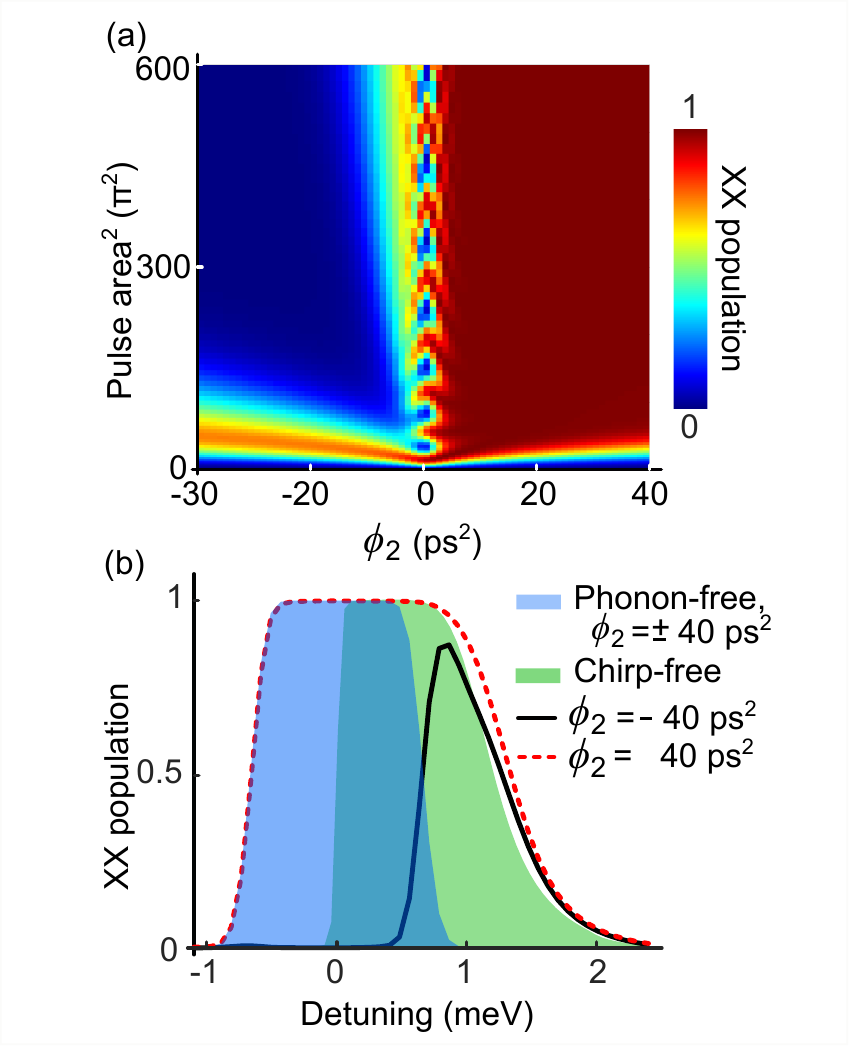}
    \caption{\textbf{Theoretical calculations}: (a) Biexciton (XX) population under ARP at $T=\SI{1}{K}$ as function of $\phi_{2}$ and pulse area, starting from a transform limited pulse of $\tau_0=\SI{2.72}{ps}$ tuned to the TPE resonance. 
    (b) Biexciton (XX) population as a function of detuning for chirped excitation ($\phi_{2}=\pm\SI{40}{\pico\second\squared}$) for $\Theta=20\pi, \tau_0=\SI{2.72}{ps}$ without phonons (blue shaded area) and including phonons at $T=\SI{1}{K}$ (black and red-dashed lines) 
    The green shaded area shows the pure phonon-assisted excitation without chirp at $\tau_0=\SI{41}{ps}$ and $\Theta=78\pi$. 
    }
    \label{fig:biexciton_chirp}
\end{figure}

The chirped laser pulses exciting the quantum dot induce the ARP process to achieve high-fidelity inversion of the biexciton state.
The ARP mechanism relies on a sweep of the instantaneous frequency of the chirped laser pulse across the quantum dot resonance (or electrical Stark tuning under a constant laser pulse \cite{mukherjee_electrically_2020}) to achieve a complete population inversion via an avoided energy level crossing in the dressed state picture \cite{hui_proposal_2008,malinovsky_general_2001}.
In quantum dots, the ARP mechanism can be applied to excite both the exciton \cite{debnath_high-fidelity_2013,simon_robust_2011,mathew_subpicosecond_2014} and the biexciton \cite{glassl_biexciton_2013,wei_deterministic_2014,kaldewey_coherent_2017} states. Because quantum dots are embedded in a solid-state environment, the interaction with the lattice vibrations (phonons) can cause a deterioration of the ARP process, which for low temperatures depends on the sign of the chirp \cite{luker_influence_2012}. 
We performed calculations in a standard three-level model consisting of ground $\ket{g}$, exciton $\ket{x}$ and biexciton state $\ket{xx}$ including the exciton-phonon interaction for longitudinal acoustic phonons \cite{kaldewey_coherent_2017,luker_phonon_2017, reiter_distinctive_2019} (for a detailed description see SI). 
With this, we obtain the biexciton generation for a chirped pulse at $T=\SI{1}{\kelvin}$ in Fig.~\ref{fig:biexciton_chirp}(a), assuming a biexciton binding energy of \SI{4}{\milli\electronvolt}. 
At $\phi_{2}= \SI{0}{ps^{2}}$ we observe Rabi rotations. For $\phi_{2}$ > 0, we obtain a large biexciton population in a large range of pulse area and chirp, while for $\phi_{2}$ < 0, phonons hinder the efficient population of the biexciton. 

In the experiments, as shown in Fig.~\ref{fig:setup}(b), the laser beam coming from the stretcher is fiber-coupled and directed to a closed-cycle cryostat (base temperature \SI{1.5}{\kelvin}, ICEOxford), where the quantum dot sample is mounted on a three-axis piezoelectric stage (ANPx101/ANPz102, attocube systems AG). A programmable electronic variable attenuator (VOA, V800PA, Thorlabs) helps sweeping the pulse power. The setup employs a cross-polarization filtering configuration for efficient laser scattering rejection. The pulse powers are monitored with a 1\% reflector [not shown in Fig.~\ref{fig:setup}(b)] before the polarizing beamsplitter. 
 
The excitation laser beam is focused onto the quantum dots with a cold objective (NA = 0.81, attocube systems AG). The quantum dot emission is collected via the same path, and the exciton (X) and biexciton (XX) photons are spectrally and spatially separated by a home-built monochromator equipped with four narrow-band notch filters (BNF-805-OD3, FWHM \SI{0.3}{\nano\meter}, Optigrate). 
The separated photons are routed to a single-photon sensitive spectrometer (Acton SP-2750, Roper Scientific) equipped with a liquid Nitrogen cooled charge-coupled device camera (Spec10 CCD, Princeton Instruments) or superconducting nanowire single-photon detectors (SNSPD, Eos, Single Quantum) for lifetime, cross-correlation and Hanbury Brown-Twiss (HBT) measurements.  

\begin{figure*}[!hbt]
    \centering
    \includegraphics[width=\linewidth]{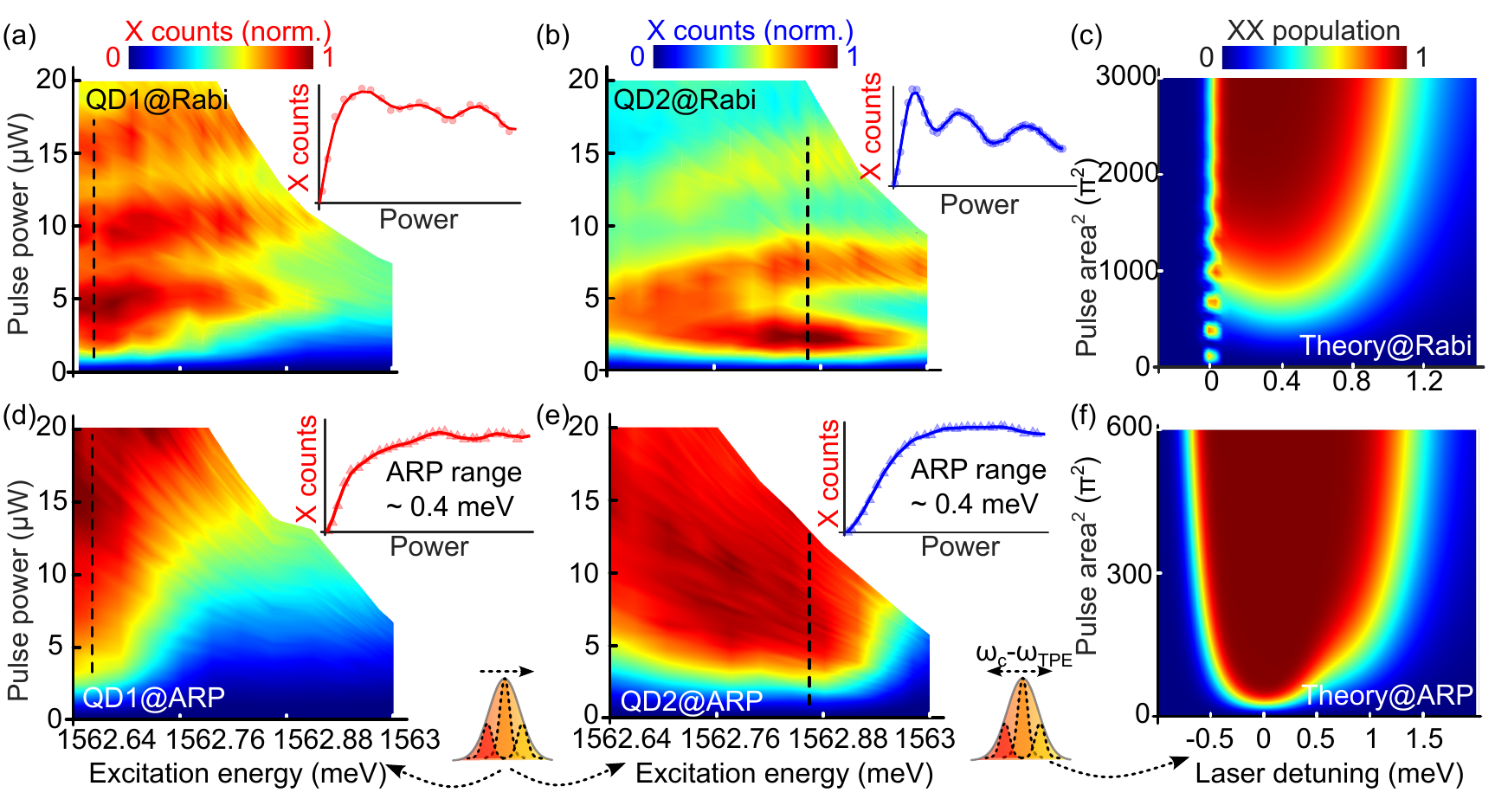}
    \caption{\textbf{Sensitivity of resonant TPE and versatility of ARP}: 
    Excitation and emission-frequency-resolved TPE Rabi measurements (no chirp, i.e., $\phi_{2} = \SI{0}{\pico\second\squared}$) represented by measured X photon counts for QD1 [panel (a)] and QD2 [panel (b)]. Corresponding results at ARP condition ($\phi_{2} = \SI{40}{\pico\second\squared}$) for QD1 [panel (d)] and QD2 [panel (e)]. The respective insets show the linecuts at respective TPE resonances (indicated by the black dashed lines on the two-dimensional maps). The integrated photon counts are normalized for QD1 and QD2 individually, but are the same at Rabi ($\phi_{2} = \SI{0}{\pico\second\squared}$) and ARP ($\phi_{2} = \SI{40}{\pico\second\squared}$) conditions. (c) Theoretical calculation of the biexciton preparation efficiency under Rabi ($\phi_{2} = \SI{0}{\pico\second\squared}$) and (f) ARP condition ($\phi_{2} = \SI{40}{\pico\second\squared}$).
    } 
    \label{fig:ARPdetuning}
\end{figure*}

\section*{Robustness for simultaneous excitation} 
The main goal of this work is to excite multiple quantum dots simultaneously. On the single quantum dot level, as evident in Fig.~\ref{fig:biexciton_chirp}(a), ARP is robust against fluctuations in pulse area and chirp. 
To excite spectrally distinct quantum dots, we further investigate the efficiency of the biexciton occupation as a function of detuning as shown in Fig.~\ref{fig:biexciton_chirp}(b) for a chirp of $|\phi_2|=\SI{40}{\pico\second\squared}$ and a pulse area of $\Theta = \SI{20}{\pi}$. The detuning $\Delta$ is defined as difference of the central laser frequency to the TPE resonance, i.e., $\Delta= \hbar(\omega_c-\omega_{\text{TPE}})$. 

Without phonons [blue area in Fig.~\ref{fig:ARPdetuning}(b)], the biexciton occupation does not depend on the sign of $\Delta$ and we find that for detunings for a detuning range of $\SI{-0.45}{\milli\electronvolt} \lesssim \Delta \lesssim  \SI{0.45}{\milli\electronvolt}$, the biexciton state still gets fully occupied, relying on the ARP process. 

If we now include phonons, the behaviour of the final occupation changes drastically. For negative chirp [black curve in Fig.~\ref{fig:ARPdetuning}(b)], phonons destroy the ARP process [see also Fig.~\ref{fig:biexciton_chirp}(a)], and the laser pulses do not lead to any biexciton occupation for a detuning range of $\SI{-0.45}{\milli\electronvolt} \lesssim \Delta \lesssim  \SI{0.45}{\milli\electronvolt}$, while for $|\Delta|\approx \SI{0.8}{\milli\electronvolt}$, the biexciton state can get $\approx$80\% occupied, due to phonon-assisted processes (for details and an explanation in the dressed states, see SI). 

Most interestingly, for positive chirp [red-dashed curve in Fig.~\ref{fig:ARPdetuning}(b)] the phonons do not hinder the ARP process. Instead, they expand the detuning range suitable for high fidelity preparation ($\geq$95\%) significantly: $\SI{-0.45}{\milli\electronvolt} \lesssim \Delta \lesssim \SI{0.75}{\milli\electronvolt}$. We refer to this widening of the detuning window as \emph{phonon advantage}. This implies that, in principle, using chirped pulses, one can excite biexciton states that are detuned as large as \SI{1.2}{\milli\electronvolt} in a quantum dot ensemble.

To underline the effects of the phonon-assisted preparation further, we compare the results to the chirp-free scenario [$\phi_{2}=\SI{0}{\pico\second\squared}$, shown as green area in Fig.~\ref{fig:biexciton_chirp}(b)], but with a much higher pulse area of $\SI{78}{\pi}$ and $\tau_0=\SI{24}{ps}$ (while for the ARP calculation we have used a pulse area of $\SI{20}{\pi}$, which corresponds to the same pulse shape after chirping, see SI). 
For positive detuning $\Delta>0$, the phonon-assisted process leads to a high biexciton occupation, which agrees with the occupation in the chirped case. 

In summary, we observe that, for chirped excitation, two different mechanisms lead to the biexciton occupation for positive detuning: the ARP process (for $\abs{\Delta} \lesssim  \SI{0.45}{\milli\electronvolt}$) and the phonon-assisted effect (for $\SI{0.45}{\milli\electronvolt} \lesssim \Delta \lesssim  \SI{0.75}{\milli\electronvolt}$) (see SI for a detailed discussion).

Given the theoretical background, we now search for a pair of quantum dots, which are spectrally separated within a window of \SI{1}{\milli\electronvolt}. Our sample consists of GaAs-AlGaAs quantum dots with exciton emission centered around \SI{790}{\nano\meter} grown by the Al-droplet etching method \cite{huber2017highly,da2021gaas}. 

We locate two bright quantum dots that are spatially separated by $\SI{1}{\micro\meter}$, labelled QD1 (red) and QD2 (blue), [see Fig.~\ref{fig:ARPdetuning}(c)], having their characteristic exciton emission lines at $X_{\text{QD1}}=\SI{792.49}{\nano\meter}$ and $X_{\text{QD2}}=\SI{792.37}{\nano\meter}$, i.e., they are spectrally separated by \SI{0.12}{\nano\meter}, or \SI{0.2}{\milli\electronvolt}. These quantum dots are not expected to dipole-couple to each other over such a distance. The corresponding TPE spectra, individually measured, are presented in Fig \ref{fig:setup} (c), alongside their representative energy level schemes. 

\section*{Individual excitation}

We start with the characterization of the two quantum dots QD1 and QD2 [see Fig.~\ref{fig:setup}(c)] and investigate the robustness of the ARP in comparison with the Rabi rotations. To quantitatively illustrate the sensitivity of Rabi rotations to the excitation conditions and quantum dot inhomogeneous broadening, we perform the TPE experiment on QD1 and QD2. The excitation- and emission-frequency resolved results are displayed in Fig.~\ref{fig:ARPdetuning}(a) and (b). 
Here, the central laser frequency is scanned in 64 steps from \SI{1562.6}{\milli\electronvolt} to \SI{1563}{\milli\electronvolt}, exploiting the motorized slit in the Fourier plane of the grating stretcher. 
For every position of the motorized slit, i.e., for a fixed central laser frequency, the TPE Rabi experiment is performed, by sweeping the pulse power and recording the TPE emission spectra for QD1 and QD2, after which the photon counts at X and XX emission energies are integrated to prepare the spectrally resolved maps (for brevity, only the photon counts at X emission energy are presented here). We note that the contribution of phonon-assisted TPE is negligible in the explored range of parameters.
The linecuts (represented by black dashed lines on the two dimensional maps) denote the respective TPE resonance conditions of QD1 and QD2. 
The photon counts are normalized to the individual maxima for QD1 and QD2. 

For the spectral range \SI{1562.6}{\milli\electronvolt} to \SI{1563}{\milli\electronvolt} considered in the experiment, we observe a clear distinction in the X emission landscapes of QD1 and QD2 [Fig.~\ref{fig:ARPdetuning} (a) and (b)]. For QD1, the resonance frequency is observed at $\approx \SI{1562.64}{\milli\electronvolt}$ and $\pi$ power is found to be $\approx$ \SI{5}{\micro\watt}, while for QD2 the resonance is found at $\approx \SI{1562.80}{\milli\electronvolt}$ with a $\pi$ power of $\approx$ \SI{2.5}{\micro\watt}. 
In other words, the optimal excitation conditions for QD1 will fail to achieve more than 60\% population of the biexciton state in QD2 (for example at $\approx$ \SI{1562.64}{\milli\electronvolt}). 
This clearly demonstrates the deficiency of Rabi rotation as a universal excitation scheme for multi-quantum dot photon sources.  

In Fig.~\ref{fig:ARPdetuning}(c) we show the theoretical calculation of the biexciton occupation under TPE. For low pulse powers, it is clear that, as soon as the laser frequency shifts by \SI{0.01}{\milli\electronvolt}, the biexciton occupation drops rapidly to just 1\%, asserting its sensitivity to the detuning. Only for high pulse areas $\approx \SI{45}{\pi}$, much beyond the power used in the experiment, would one benefit from the phonon-assisted preparation scheme, that provides a high occupation for the same detuning. 

We now turn our attention to the applicability of ARP against excitation spectral shifts and quantum dot spectral shifts. For this we fix $\phi_{2}\approx\SI{40}{\pico\second\squared}$ and perform the same experiment on QD1 and QD2 [Fig.~\ref{fig:ARPdetuning}(d) and (e)]. 
The resulting population landscape is largely a plateau for QD1 and QD2, despite an excitation frequency scan of \SI{0.4}{\milli\electronvolt}, relying on the ARP process. 
For QD1, the exciton photon counts largely remain stable after \SI{15}{\micro\watt} onwards, while for QD1, plateau is achieved from \SI{10}{\micro\watt} onwards, which reflect the differences in their $\pi$ pulse powers. 

Therefore, it is clear that despite the energetic separation, through ARP, both QD1 and QD2 can be excited to the biexciton state with high fidelity within an excitation frequency scan range of \SI{0.4}{\milli\electronvolt}. 

The robustness of the preparation is again confirmed in the theoretical calculations in Fig.~\ref{fig:ARPdetuning}(f) showing an excellent agreement with experiment. It also clearly demonstrates the region of phonon advantage that is asymmetric with respect to the sign of detuning as discussed in Fig.~\ref{fig:biexciton_chirp}. 

Furthermore, we also investigated the single-photon quality via HBT measurements and obtained the second-order photon correlation of QD2 at Rabi condition as $g^{(2)}(0)_{\text{QD1}}=0.02$ and at ARP regime as $g^{(2)}(0)_{\text{QD2}}=0.05$, respectively, asserting that the single-photon characteristics are maintained at both regimes (see SI). 

\section*{Simultaneous excitation of two quantum dots}
\begin{figure}[!hbt]
    \centering
    \includegraphics[width=\linewidth]{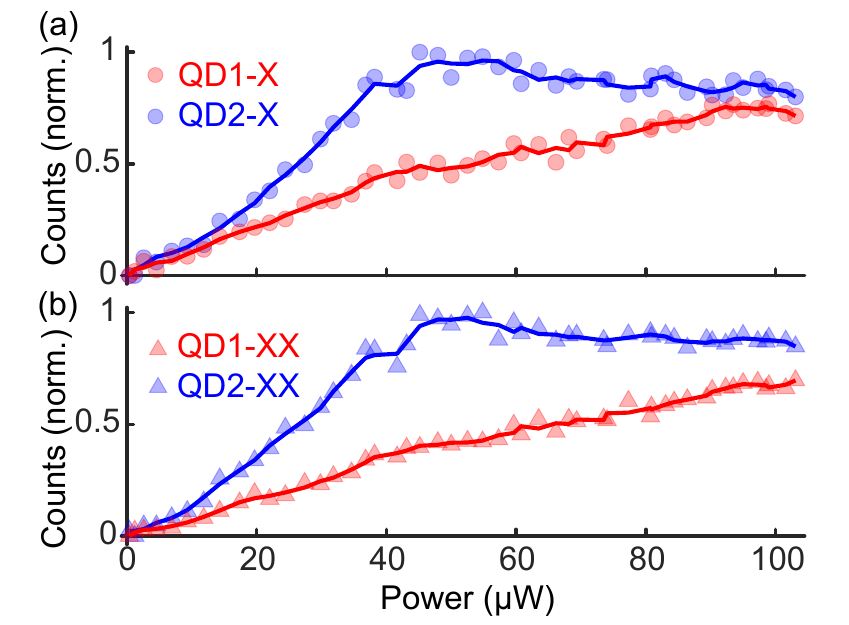}
    \caption{\textbf{Collective excitation at ARP}: Integrated photon counts at (a) X and (b) XX emission energies for QD1 and QD2 recorded under simultaneous excitation, at $\phi_{2} = \SI{40}{\pico\second\squared}$ 
    by positioning the excitation laser spot between QD1 and QD2. The photon counts counts are normalized to a common maximum, respectively for X and XX emissions.
    }
    \label{fig:ARPdot1and2}
\end{figure}

Lastly, we investigated the versatility of ARP to collectively excite QD1 and QD2. For this, after setting $\phi_{2}\approx\SI{40}{\pico\second\squared}$, we position the sample so as to simultaneously excite and collect from both QD1 and QD2, keeping the objective fixed and the frequency spectrum centered at \SI{1562.6}{\milli\electronvolt}. We then perform a laser power sweep, recording the collective ARP spectra with a single-mode fiber coupler. 
In Fig.~\ref{fig:ARPdot1and2} we display the integrated photon counts at X and XX emission energies for both quantum dots, normalized to the maximum in either case. We observe that the photon counts of both quantum dots achieve plateaus at high pulse areas (i.e., at ARP regime). This reflects the response we have witnessed earlier in Fig.~\ref{fig:ARPdetuning}. 

Thus, we have successfully produced two spatially separated quantum dot photon sources that are collectively excited, despite their energetic separation, without modifying the excitation parameters. This is remarkable: a simple excitation scheme with displaced beam can achieve a simultaneous and high efficiency preparation of two distinct biexciton states, in two spatially and spectrally different quantum dots. 
Our scheme is much simpler than an alternative, dual-confocal excitation scheme \cite{duquennoy_real-time_2022}, which, considering the pulse power requirement at high chirp ($\phi_{2} = \SI{40}{\pico\second\squared}$), is challenging, especially when trying to scale it to more than two quantum dots. On the other hand, in nanowire quantum dots, where individual quantum dots are separated along the growth direction \cite{koong_multiplexed_2020}, one could relatively easily implement our method to upscale the production of entangled photons. 

\section*{Conclusions}

To conclude, we have presented that simultaneous, high-fidelity preparation of biexciton states is possible in multiple, spectrally distinct quantum dots via chirped laser excitation, relying on adiabatic rapid passage. 
Initially, we showed that near-unity preparation efficiency of biexciton states is maintained in the chosen quantum dots for an excitation energy shift as large as \SI{0.4}{\milli\electronvolt}, clearly establishing the versatility of this scheme against laser frequency fluctuations and quantum dot spectral detunings. 
Furthermore, we have validated the robustness of the adiabatic scheme by simultaneously exciting the two quantum dots to near-unity biexciton population.
We have also presented the theoretical modeling of the scheme based on dressed states, and determined a regime of phonon advantage, whereby the spectral detuning range can be greatly expanded in addition to reducing the pulse power requirements.  
Our scheme can be generalized to any quantum dot system and is a significant contribution towards a robust excitation method to scale-up quantum dot based quantum photonic architecture.

\section*{Acknowledgements}
The authors acknowledge Julian Münzberg for insightful discussions. FK, YK, VR and GW acknowledge financial support through the Austrian Science Fund FWF projects W1259 (DK-ALM Atoms, Light, and Molecules), FG 5, TAI-556N (DarkEneT) and I4380 (AEQuDot). TKB and DER acknowledge financial support from the German Research Foundation DFG through the project 428026575 (AEQuDot). AR and SFCdS acknowledge C. Schimpf for fruitful discussions, the FWF projects FG 5, P 30459, I 4320,  the Linz Institute of Technology (LIT) and the European Union's Horizon 2020 research, and innovation program under Grant Agreement Nos. 899814 (Qurope), 871130 (ASCENT+).


\clearpage
\appendix
\section{Supporting Information}
\subsection{Theoretical model}
The quantum dot is approximated by a three-level system consisting of ground state $\ket{g}$, exciton state $\ket{x}$ as well as a biexciton state $\ket{xx}$. The exciton and biexciton are addressed by a linearly polarized, pulsed laser $\Omega_X(t)$, described in dipole and rotating wave approximation. The Hamiltonian for this system reads
\begin{align}
\begin{split}
    H = &\hbar\omega_0\ket{x}\bra{x} +\hbar(2\omega_0-\Delta_B)\ket{xx}\bra{xx} \\
&- \frac{\hbar}{2}\left[\Omega_X(t)\left(\ket{g}\bra{x}+\ket{x}\bra{xx}\right)+h.c.\right]. 
\end{split}\label{eq:hamiltonian_cl}
\end{align}
While a single exciton is separated from the ground state by the energy $\hbar\omega_0$, the biexciton exhibits the so-called biexciton binding energy $\hbar\Delta_B$, such that two-photon processes resonant to $\omega_0-\Delta_B/2$ are possible. For the dots considered here, $\hbar\Delta_B$ amounts to $\SI{4}{meV}$.
To describe the chirping process, we consider a Gaussian transform-limited pulse  
\begin{equation}
    \Omega_{\text{0}}(t) = \frac{\Theta}{\sqrt{2\pi}\sigma_0}e^{-\frac{t^2}{2\sigma_0^2}}e^{-i\omega_c t},
\end{equation}
with the pulse area $\Theta$, the frequency $\omega_c$ and the pulse duration $\sigma_0$. The action of the chirper results in a phase of the pulse  
\begin{equation}
    \phi(t)=\omega_c t + \frac{a}{2}t^2,
\end{equation}
with the temporal chirp $a=\frac{\phi_2}{\phi_2^2+\sigma_0^2}$ determined by the GDD $\phi_2$.
That leads to a pulse of the form
\begin{equation}
    \Omega_{X}(t) = \frac{\Theta}{\sqrt{2\pi\sigma\sigma_0}}e^{-\frac{t^2}{2\sigma^2}}e^{-i\phi(t)}.
\end{equation}
Note that the effective pulse area of the chirped pulse is modified due to the chirping process.

The chirping is accompanied by a change of the pulse duration according to $\sigma=\sqrt{(\phi_2/\sigma_0)^2+\sigma_0^2}$ \cite{malinovsky_general_2001}. In the theoretical calculation we consider the pulse duration of the electric field. In the experimental data, the pulse duration $\tau$ corresponds to the intensity full width of half maximum (FWHM). Hence, these quantities are connected via
\begin{equation}
    \tau_p = 2\sqrt{\ln2}\sigma \,.
\end{equation}

The detuning between the laser pulse and the three-level system is defined such that resonant excitation corresponds to the two-photon excitation of the biexciton, i.e.,
\begin{equation}
    \Delta = \omega_c-\omega_{\text{TPE}}=\omega_c-(\omega_0-\Delta_B/2).
\end{equation}

For the interaction with the phonon environment stemming from the bulk material which the quantum dot is embedded in, we use the standard pure-dephasing Hamiltonian describing the coupling to longitudinal acoustic phonons, such that the additional phonon part reads
\begin{align}
\begin{split}
        H_{\text{ph}} &= \hbar\sum_{\q}\omega_{\q}^{}b^{\dagger}_{\q}b_{\q}^{}\\
        &+ \hbar\left(\ket{x}\bra{x} + 2\ket{xx}\bra{xx}\right)\sum_{\q}\left(g_{\q}^{}b_{\q}^{}+g_{\q}^{*}b_{\q}^{\dagger}\right).
\end{split}
\label{eq:hamiltonian_phonons}
\end{align}
Here, $b^{}_{\q}(b^{\dagger}_{\q})$ is a bosonic operator destroying (creating) a phonon with wave vector $\q$ and corresponding energy $\hbar\omega_{\q}$. Coupling between phonons and exciton states is mediated by the coupling element $g_{\q}^{}$. We consider coupling to longitudinal acoustic phonons with linear dispersion and deformation potential coupling. We take the material parameters from Ref. \cite{kaldewey_coherent_2017} (in the SI).

Since the processes depend on the number of excitons that are present in the system, the phonon interaction is twice as strong for the biexciton state involving two exciton states. Also, the energy shift due to the formation of a polaron is different for exciton and biexciton states. As it depends on the square of the coupling, it is four times stronger for the biexciton, compared to the exciton. To compare calculations with and without phonons, in the case including phonons, the polaron energy shift is subtracted from the system energies, so that the energy difference between exciton and biexciton again equals the biexciton binding.

In order to quantify the strength of the carrier-phonon interaction, a useful measure is the phonon spectral density
\begin{equation}
    J(\omega) = \sum_{\q}|g_{\q}|^2\delta(\omega-\omega_{\q}).
\end{equation}

This quantity is highly dependent on the size of the dot, while for the electronic properties only, the specific shape of the dot is not important \cite{luker_phonon_2017}. We assume a $\SI{5}{nm}$ spherical GaAs quantum dot, which has its maximum of the phonon spectral density at approximately $\SI{1.2}{\milli\electronvolt}$.

The equations of motions are then calculated via the von-Neumann equation and the Hamiltonian in Eq.~\eqref{eq:hamiltonian_phonons}, which leads to an infinite hierarchy of differential equations due to the unrestricted dimension of the resulting Hilbert space. We truncate this hierarchy using a correlation expansion approach. Neglecting processes involving three or more phonons, corresponding to a fourth order Born approximation, has been shown to result in an adequate description of this system \cite{kaldewey_coherent_2017}. In particular, it achieves good agreement with a numerically-exact path-integral formalism \cite{glassl_long-time_2011}.

\subsection{Dressed-state picture}
\begin{figure}[t]
    \centering
    \includegraphics[width=\linewidth]{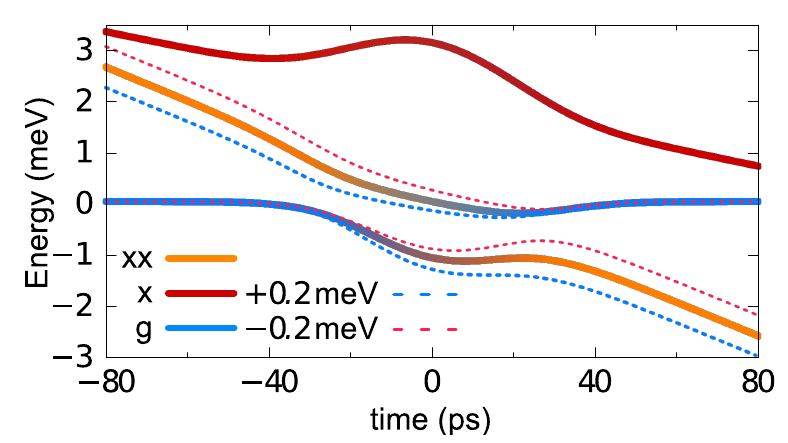}
    \caption{
    \textbf{ARP in the dressed-state picture:}
    Dressed state energies during a pulse with $\Theta=20\pi, \phi_2=\SI{40}{\pico\second\squared}, \sigma_0=\SI{1.62}{ps}$, corresponding to $\tau=\SI{2.72}{ps}$. Solid lines correspond to detuning $\Delta=0$ 
    The changing color of the line reflects the composition of the respective dressed states from the bare states. 
    The dashed blue (red) lines are with a detuning of $+\SI{0.2}{meV}(\SI{-0.2}{meV})$ (only for $g$ and $xx$). 
    }
    \label{fig:dressed_states_arp}
\end{figure}

To understand the processes leading to ARP and phonon assistance, an illustrative picture is provided by dressed states. For this, the Hamiltonian in Eq.~\eqref{eq:hamiltonian_cl}, including the carrier-light interaction but not the phonon interaction, is transformed into a reference frame rotating with the laser frequency. By doing this, the coupling terms $\Omega_X$ become real quantities. The transformed Hamiltonian is then diagonalized, leading to the dressed-state energies and dressed eigenstates of the system. 

Figure~\ref{fig:dressed_states_arp} shows the time evolution of the dressed-state energies for a chirped excitation with $\phi_2=+\SI{40}{\pico\second\squared}$. The color of the lines correspond to how the dressed states are mixed from the bare states. Note that the exciton state $x$ is also present in this system, but it is rather far off and therefore is not involved in the ARP process.
Before and after the pulse, the states do not mix. During the pulse, the chirp leads to an anticrossing of the dressed-state energies and a switching of the corresponding bare states. This means, the branch that previously consisted of the biexciton (orange) now consists of the ground state (blue) and vice versa. If the evolution is adiabatic, i.e., no transitions between the dressed states happen, this results in a full transfer of the ground state to the excited state.

\begin{figure}[!t]
    \centering
    \includegraphics[width =\columnwidth]{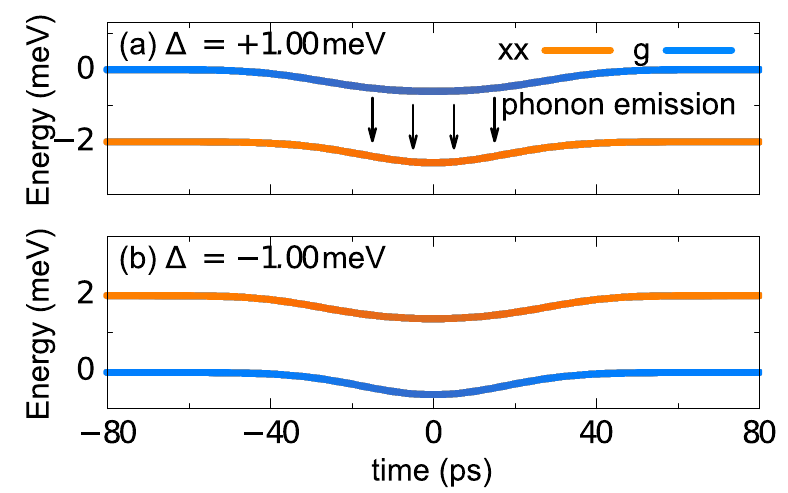}
    \caption{
    \textbf{Phonon-assisted preparation in the dressed-state picture:}
    Dressed-state energies during a pulse without chirp ($\phi_2=0$) and $\Theta=78.16\pi$ and  $\sigma_0=\SI{24.74}{ps}$.
    for (a) positive detuning of $\Delta=\SI{1.00}{meV}$ and (b) negative detuning of $\Delta=\SI{-1.00}{meV}$. The dressed state corresponding mainly to the exciton has been omitted here. During the pulse, phonon emission  can occur for positive detuning as indicated by black arrows.
    }
    \label{fig:dressed_states_unchirped}
\end{figure}

To understand the influence of phonon-assisted processes, the dressed-state energies for unchirped excitation are shown in Fig.~\ref{fig:dressed_states_unchirped}. The corresponding biexciton occupation as function of detuning and pulse power is shown in the main text in Fig.~\ref{fig:ARPdetuning}(c)(top), where we find that for positive detuning and high powers the biexciton becomes populated. In Fig.~\ref{fig:dressed_states_unchirped}(a), a positive detuning of $\SI{1.00}{meV}$ is chosen at a pulse area of $\Theta=78\pi$ and pulse during of $\sigma_0=\SI{24}{ps}$. These values correspond to the same envelope shape compared to a chirped pulse with $\phi_{2}=\SI{40}{ps^2},\Theta=20\pi,\sigma_0=\SI{1.62}{ps}$. Before and after the pulse, the upper dressed state corresponds to the ground state $\ket{g}$ and the lower to the biexciton state $\ket{xx}$. During the pulse, the system only mixes slightly and the branches do not (anti)cross like in the case of ARP. Instead, during the pulse phonon assisted processes can bridge the energy gap between the dressed states. Transitions between the dressed states are now allowed, converting ground to biexciton state under the emission of phonons. At low temperature, only phonon emission processes are possible, while phonon absorption processes cannot happen, because there are no phonons available to absorb.

Choosing a negative detuning (as shown in Fig.~\ref{fig:dressed_states_unchirped}(b)), phonon emission processes are no longer possible. Now, the upper state corresponds to the biexciton. This explains the asymmetry regrading the detuning as observed in Fig.~\ref{fig:ARPdetuning}(c,top).

It is important to note that the splitting of the dressed states is crucial for processes including phonons: The splitting has to lie within the range of the phonon spectral density for the effects to happen efficiently. If either the detuning or the pulse area is too high, the splitting is too large and phonon influence is suppressed. For resonant excitation, this leads to the so-called reappearance regime for high pulse area under resonant excitation of a two-level system.

\begin{figure}[!t]
    \centering
    \includegraphics[width =\columnwidth]{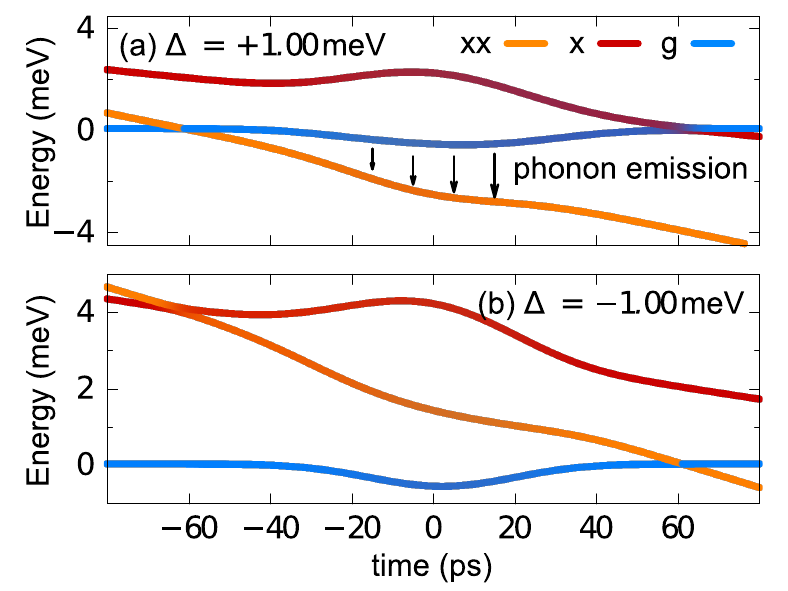}
    \caption{
    \textbf{Phonon advantage in the dressed-state picture:} Same as Fig.~\ref{fig:dressed_states_unchirped}, but now for a chirped pulsed with $\phi_2=\SI{40}{\pico\second\squared}$ (and $\Theta=20\pi, \sigma_0=\SI{1.62}{ps}$).
    }
    \label{fig:dressed_states_chirped_detuning}
\end{figure}

The most interesting case leading to the phonon advantage is when we use a chirped pulse with a large detuning. The dressed state energies for these cases are shown in Fig.~\ref{fig:dressed_states_chirped_detuning}. Now, the detunings are so large that the dressed-state energies no longer show an anticrossing, but cross at $t\approx\SI{60}{ps}$ before (a) or at $t\approx\SI{-60}{ps}$ after (b) the pulse, depending on the sign of the detuning. No ARP process can take place for these parameters, so without phonons no excitation is possible (cf. Fig.~\ref{fig:biexciton_chirp}(b)). Nevertheless, the splitting of the dressed state energies again allows for phonon assisted processes. For a positive detuning of $\SI{1.00}{meV}$ shown in Fig.~\ref{fig:dressed_states_chirped_detuning}(a), as usual the ground state corresponds to the upper dressed state, so the biexciton can be prepared under phonon emission. In Fig.~\ref{fig:dressed_states_chirped_detuning}(b), a negative detuning of $\SI{-1.00}{meV}$ is used. Now the biexciton corresponds to the upper dressed state such that phonon absorption would be required.
In both cases, the dressed state branch corresponding to the exciton $x$ intersects with the $g$ or $xx$ branch, but at this point no population is transferred, because the laser pulse is not active yet. 

The dressed state picture can also be used to understand the influence of phonons on the excitation using a negatively chirped pulse. In the dressed state picture, changing the sign of the chirp effectively mirrors the time-axis, so the diagrams have to be read right-to-left instead. For the cases shown in Fig.~\ref{fig:dressed_states_chirped_detuning}, this does not change the effects as in (a), still phonon emission leads to preparation of the biexciton state and in (b) no excitation is possible. However, the big difference for $\phi_2=-\SI{40}{\pico\second\squared}$ is under no or only slight detuning, the process does not work including phonons. For this, we look back at Fig.~\ref{fig:dressed_states_arp} and again read it right-to-left. Now, in the beginning the upper branch corresponds to the ground state and the lower branch to the biexciton. During the pulse, phonon emission happens from the upper to the lower branch, which after the pulse corresponds to the ground state. This explains why ARP with negative chirp is strongly disturbed by phonons.

\subsection{Photon characterization}
\begin{figure}[!hbt]
    \centering
    \includegraphics[width=1\linewidth]{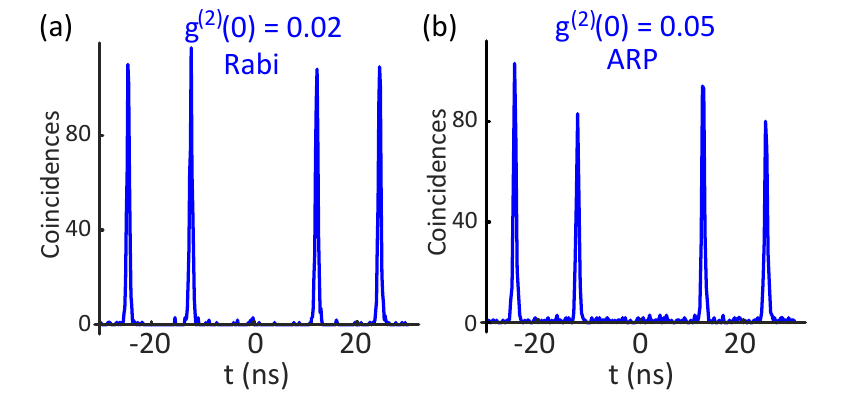}
    \caption{\textbf{Photon characterization}: (a) Measured single-photon characteristics at Rabi condition, i.e. $\phi_{2} = \SI{0}{\pico\second\squared}$ and pulse power of \SI{2.5}{\micro\watt} (b) at ARP condition, i.e. $\phi_{2} = \SI{40}{\pico\second\squared}$ and pulse power of \SI{7}{\micro\watt}.} 
    \label{fig:g2both}
\end{figure}

To verify the photon quality, we performed HBT measurements (see Fig.~\ref{fig:g2both}) for X photons for QD2 at the Rabi ($\phi_{2} = \SI{0}{\pico\second\squared}$) and the ARP $\phi_{2} = \SI{40}{\pico\second\squared}$ conditions, and observed those to be $g^{(2)}(0) = 0.02$ and $g^{(2)}(0) = 0.05$ respectively. Therefore, the single photon characteristics are preserved in both excitation regimes. The nominal increase in $g^{(2)}(0)$ for ARP condition is attributed to the laser scattering at higher pulse powers. 

\subsection{Chirp characterization}

\begin{figure}[!hbt]
    \centering
    \includegraphics[width=\linewidth]{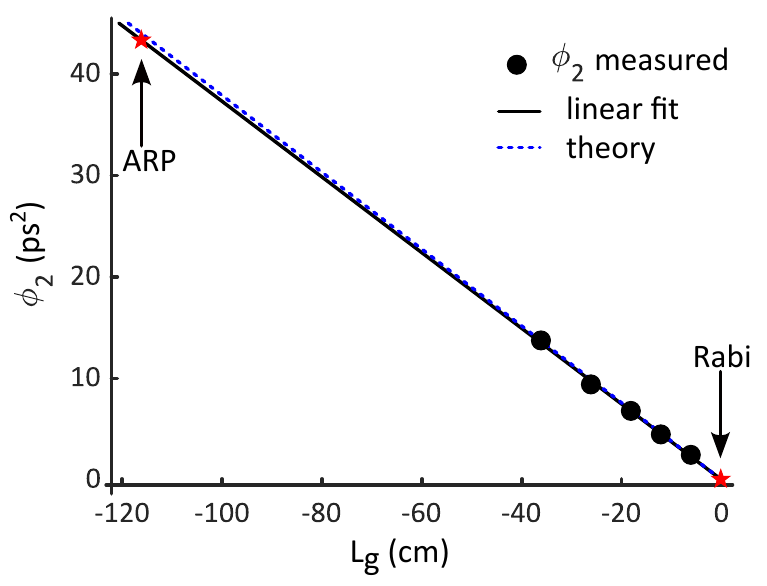}
    \caption{\textbf{Stretcher characterization}: Calibrated values of $\phi_{2}$ in the grating stretcher employed in the experiments. The solid black line is a linear fit of the measured data points (black dots) while the dashed lines in blue show the calculated value of $\phi_{2}$. Red stars indicate the $\phi_{2}$ values at which the experiments were performed.} 
    \label{fig:chirperdata}
\end{figure}

Our grating stretcher is constructed in a folded, single grating geometry \cite{martinez_3000_1987,backus_high_1998}. In Fig.~\ref{fig:chirperdata}(blue dashed line) we display the calculated GDD values for various grating positions in our grating stretcher \cite{backus_high_1998}
$$
\phi_{2} = \dv[2]{\phi_{c}(\omega)}{\omega} =\frac{-\lambda_{0}^{3} L_{g}}{\pi c^{2}d^{2}}\left[1-\left(\frac{\lambda}{d}-\sin \theta_{i}\right)^{2}\right]^{-3 / 2}
$$
where $\lambda_{0} = \SI{793}{\nano\meter}$ is the central wavelength, $L_{g} = -2(f-s)$ is the effective grating distance (for double pass, with $f$ being the focal length of the lens and $s$ being the distance between the lens and the grating), $d = \SI{1200}{\per\milli\meter}$ is the grating period and $\theta_{i} = \SI{2}{\degree}$ is the angle of incidence. We performed the ARP experiments at $s = \SI{20}{\centi\meter}$, providing a $\phi_{2}\approx \SI{40}{ps^{2}}$ .
The red stars in Fig.~\ref{fig:chirperdata} indicate the values of $\phi_2$ during the experiments.

\clearpage
\bibliography{references.bib}%

\begin{thebibliography}{56}%
\makeatletter
\providecommand \@ifxundefined [1]{%
 \@ifx{#1\undefined}
}%
\providecommand \@ifnum [1]{%
 \ifnum #1\expandafter \@firstoftwo
 \else \expandafter \@secondoftwo
 \fi
}%
\providecommand \@ifx [1]{%
 \ifx #1\expandafter \@firstoftwo
 \else \expandafter \@secondoftwo
 \fi
}%
\providecommand \natexlab [1]{#1}%
\providecommand \enquote  [1]{``#1''}%
\providecommand \bibnamefont  [1]{#1}%
\providecommand \bibfnamefont [1]{#1}%
\providecommand \citenamefont [1]{#1}%
\providecommand \href@noop [0]{\@secondoftwo}%
\providecommand \href [0]{\begingroup \@sanitize@url \@href}%
\providecommand \@href[1]{\@@startlink{#1}\@@href}%
\providecommand \@@href[1]{\endgroup#1\@@endlink}%
\providecommand \@sanitize@url [0]{\catcode `\\12\catcode `\$12\catcode
  `\&12\catcode `\#12\catcode `\^12\catcode `\_12\catcode `\%12\relax}%
\providecommand \@@startlink[1]{}%
\providecommand \@@endlink[0]{}%
\providecommand \url  [0]{\begingroup\@sanitize@url \@url }%
\providecommand \@url [1]{\endgroup\@href {#1}{\urlprefix }}%
\providecommand \urlprefix  [0]{URL }%
\providecommand \Eprint [0]{\href }%
\providecommand \doibase [0]{https://doi.org/}%
\providecommand \selectlanguage [0]{\@gobble}%
\providecommand \bibinfo  [0]{\@secondoftwo}%
\providecommand \bibfield  [0]{\@secondoftwo}%
\providecommand \translation [1]{[#1]}%
\providecommand \BibitemOpen [0]{}%
\providecommand \bibitemStop [0]{}%
\providecommand \bibitemNoStop [0]{.\EOS\space}%
\providecommand \EOS [0]{\spacefactor3000\relax}%
\providecommand \BibitemShut  [1]{\csname bibitem#1\endcsname}%
\let\auto@bib@innerbib\@empty
\bibitem [{\citenamefont {Benson}\ \emph {et~al.}(2000)\citenamefont {Benson},
  \citenamefont {Santori}, \citenamefont {Pelton},\ and\ \citenamefont
  {Yamamoto}}]{benson_regulated_2000}%
  \BibitemOpen
  \bibfield  {author} {\bibinfo {author} {\bibfnamefont {O.}~\bibnamefont
  {Benson}}, \bibinfo {author} {\bibfnamefont {C.}~\bibnamefont {Santori}},
  \bibinfo {author} {\bibfnamefont {M.}~\bibnamefont {Pelton}},\ and\ \bibinfo
  {author} {\bibfnamefont {Y.}~\bibnamefont {Yamamoto}},\ }\bibfield  {title}
  {\bibinfo {title} {Regulated and {Entangled} {Photons} from a {Single}
  {Quantum} {Dot}},\ }\href {https://doi.org/10.1103/PhysRevLett.84.2513}
  {\bibfield  {journal} {\bibinfo  {journal} {Phys. Rev. Lett.}\ }\textbf
  {\bibinfo {volume} {84}},\ \bibinfo {pages} {2513} (\bibinfo {year}
  {2000})}\BibitemShut {NoStop}%
\bibitem [{\citenamefont {Akopian}\ \emph {et~al.}(2006)\citenamefont
  {Akopian}, \citenamefont {Lindner}, \citenamefont {Poem}, \citenamefont
  {Berlatzky}, \citenamefont {Avron}, \citenamefont {Gershoni}, \citenamefont
  {Gerardot},\ and\ \citenamefont {Petroff}}]{akopian_entangled_2006}%
  \BibitemOpen
  \bibfield  {author} {\bibinfo {author} {\bibfnamefont {N.}~\bibnamefont
  {Akopian}}, \bibinfo {author} {\bibfnamefont {N.~H.}\ \bibnamefont
  {Lindner}}, \bibinfo {author} {\bibfnamefont {E.}~\bibnamefont {Poem}},
  \bibinfo {author} {\bibfnamefont {Y.}~\bibnamefont {Berlatzky}}, \bibinfo
  {author} {\bibfnamefont {J.}~\bibnamefont {Avron}}, \bibinfo {author}
  {\bibfnamefont {D.}~\bibnamefont {Gershoni}}, \bibinfo {author}
  {\bibfnamefont {B.~D.}\ \bibnamefont {Gerardot}},\ and\ \bibinfo {author}
  {\bibfnamefont {P.~M.}\ \bibnamefont {Petroff}},\ }\bibfield  {title}
  {\bibinfo {title} {Entangled {Photon} {Pairs} from {Semiconductor} {Quantum}
  {Dots}},\ }\href {https://doi.org/10.1103/PhysRevLett.96.130501} {\bibfield
  {journal} {\bibinfo  {journal} {Phys. Rev. Lett.}\ }\textbf {\bibinfo
  {volume} {96}},\ \bibinfo {pages} {130501} (\bibinfo {year}
  {2006})}\BibitemShut {NoStop}%
\bibitem [{\citenamefont {Müller}\ \emph {et~al.}(2014)\citenamefont
  {Müller}, \citenamefont {Bounouar}, \citenamefont {Jöns}, \citenamefont
  {Glässl},\ and\ \citenamefont {Michler}}]{muller_-demand_2014}%
  \BibitemOpen
  \bibfield  {author} {\bibinfo {author} {\bibfnamefont {M.}~\bibnamefont
  {Müller}}, \bibinfo {author} {\bibfnamefont {S.}~\bibnamefont {Bounouar}},
  \bibinfo {author} {\bibfnamefont {K.~D.}\ \bibnamefont {Jöns}}, \bibinfo
  {author} {\bibfnamefont {M.}~\bibnamefont {Glässl}},\ and\ \bibinfo {author}
  {\bibfnamefont {P.}~\bibnamefont {Michler}},\ }\bibfield  {title} {\bibinfo
  {title} {On-demand generation of indistinguishable polarization-entangled
  photon pairs},\ }\href {https://doi.org/10.1038/nphoton.2013.377} {\bibfield
  {journal} {\bibinfo  {journal} {Nat. Photonics}\ }\textbf {\bibinfo {volume}
  {8}},\ \bibinfo {pages} {224} (\bibinfo {year} {2014})}\BibitemShut {NoStop}%
\bibitem [{\citenamefont {Basso~Basset}\ \emph {et~al.}(2019)\citenamefont
  {Basso~Basset}, \citenamefont {Rota}, \citenamefont {Schimpf}, \citenamefont
  {Tedeschi}, \citenamefont {Zeuner}, \citenamefont {Covre~da Silva},
  \citenamefont {Reindl}, \citenamefont {Zwiller}, \citenamefont {Jöns},
  \citenamefont {Rastelli},\ and\ \citenamefont
  {Trotta}}]{basso_basset_entanglement_2019}%
  \BibitemOpen
  \bibfield  {author} {\bibinfo {author} {\bibfnamefont {F.}~\bibnamefont
  {Basso~Basset}}, \bibinfo {author} {\bibfnamefont {M.}~\bibnamefont {Rota}},
  \bibinfo {author} {\bibfnamefont {C.}~\bibnamefont {Schimpf}}, \bibinfo
  {author} {\bibfnamefont {D.}~\bibnamefont {Tedeschi}}, \bibinfo {author}
  {\bibfnamefont {K.}~\bibnamefont {Zeuner}}, \bibinfo {author} {\bibfnamefont
  {S.}~\bibnamefont {Covre~da Silva}}, \bibinfo {author} {\bibfnamefont
  {M.}~\bibnamefont {Reindl}}, \bibinfo {author} {\bibfnamefont
  {V.}~\bibnamefont {Zwiller}}, \bibinfo {author} {\bibfnamefont
  {K.}~\bibnamefont {Jöns}}, \bibinfo {author} {\bibfnamefont
  {A.}~\bibnamefont {Rastelli}},\ and\ \bibinfo {author} {\bibfnamefont
  {R.}~\bibnamefont {Trotta}},\ }\bibfield  {title} {\bibinfo {title}
  {Entanglement {Swapping} with {Photons} {Generated} on {Demand} by a
  {Quantum} {Dot}},\ }\href {https://doi.org/10.1103/PhysRevLett.123.160501}
  {\bibfield  {journal} {\bibinfo  {journal} {Phys. Rev. Lett.}\ }\textbf
  {\bibinfo {volume} {123}},\ \bibinfo {pages} {160501} (\bibinfo {year}
  {2019})}\BibitemShut {NoStop}%
\bibitem [{\citenamefont {Schimpf}\ \emph
  {et~al.}(2021{\natexlab{a}})\citenamefont {Schimpf}, \citenamefont {Manna},
  \citenamefont {Da~Silva}, \citenamefont {Aigner},\ and\ \citenamefont
  {Rastelli}}]{schimpf2021entanglement}%
  \BibitemOpen
  \bibfield  {author} {\bibinfo {author} {\bibfnamefont {C.}~\bibnamefont
  {Schimpf}}, \bibinfo {author} {\bibfnamefont {S.}~\bibnamefont {Manna}},
  \bibinfo {author} {\bibfnamefont {S.~F.~C.}\ \bibnamefont {Da~Silva}},
  \bibinfo {author} {\bibfnamefont {M.}~\bibnamefont {Aigner}},\ and\ \bibinfo
  {author} {\bibfnamefont {A.}~\bibnamefont {Rastelli}},\ }\bibfield  {title}
  {\bibinfo {title} {Entanglement-based quantum key distribution with a
  blinking-free quantum dot operated at a temperature up to 20 k},\ }\href
  {https://doi.org/10.1117/1.AP.3.6.065001} {\bibfield  {journal} {\bibinfo
  {journal} {Adv. Photonics}\ }\textbf {\bibinfo {volume} {3}},\ \bibinfo
  {pages} {065001} (\bibinfo {year} {2021}{\natexlab{a}})}\BibitemShut
  {NoStop}%
\bibitem [{\citenamefont {Jayakumar}\ \emph {et~al.}(2014)\citenamefont
  {Jayakumar}, \citenamefont {Predojević}, \citenamefont {Kauten},
  \citenamefont {Huber}, \citenamefont {Solomon},\ and\ \citenamefont
  {Weihs}}]{jayakumar_time-bin_2014}%
  \BibitemOpen
  \bibfield  {author} {\bibinfo {author} {\bibfnamefont {H.}~\bibnamefont
  {Jayakumar}}, \bibinfo {author} {\bibfnamefont {A.}~\bibnamefont
  {Predojević}}, \bibinfo {author} {\bibfnamefont {T.}~\bibnamefont {Kauten}},
  \bibinfo {author} {\bibfnamefont {T.}~\bibnamefont {Huber}}, \bibinfo
  {author} {\bibfnamefont {G.~S.}\ \bibnamefont {Solomon}},\ and\ \bibinfo
  {author} {\bibfnamefont {G.}~\bibnamefont {Weihs}},\ }\bibfield  {title}
  {\bibinfo {title} {Time-bin entangled photons from a quantum dot},\ }\href
  {https://doi.org/10.1038/ncomms5251} {\bibfield  {journal} {\bibinfo
  {journal} {Nat. Commun.}\ }\textbf {\bibinfo {volume} {5}},\ \bibinfo {pages}
  {4251} (\bibinfo {year} {2014})}\BibitemShut {NoStop}%
\bibitem [{\citenamefont {Lee}\ \emph {et~al.}(2019)\citenamefont {Lee},
  \citenamefont {Villa}, \citenamefont {Bennett}, \citenamefont {Stevenson},
  \citenamefont {Ellis}, \citenamefont {Farrer}, \citenamefont {Ritchie},\ and\
  \citenamefont {Shields}}]{lee_quantum_2019}%
  \BibitemOpen
  \bibfield  {author} {\bibinfo {author} {\bibfnamefont {J.~P.}\ \bibnamefont
  {Lee}}, \bibinfo {author} {\bibfnamefont {B.}~\bibnamefont {Villa}}, \bibinfo
  {author} {\bibfnamefont {A.~J.}\ \bibnamefont {Bennett}}, \bibinfo {author}
  {\bibfnamefont {R.~M.}\ \bibnamefont {Stevenson}}, \bibinfo {author}
  {\bibfnamefont {D.~J.~P.}\ \bibnamefont {Ellis}}, \bibinfo {author}
  {\bibfnamefont {I.}~\bibnamefont {Farrer}}, \bibinfo {author} {\bibfnamefont
  {D.~A.}\ \bibnamefont {Ritchie}},\ and\ \bibinfo {author} {\bibfnamefont
  {A.~J.}\ \bibnamefont {Shields}},\ }\bibfield  {title} {\bibinfo {title} {A
  quantum dot as a source of time-bin entangled multi-photon states},\ }\href
  {https://doi.org/10.1088/2058-9565/ab0a9b} {\bibfield  {journal} {\bibinfo
  {journal} {Quantum Sci. Technol.}\ }\textbf {\bibinfo {volume} {4}},\
  \bibinfo {pages} {025011} (\bibinfo {year} {2019})}\BibitemShut {NoStop}%
\bibitem [{\citenamefont {Weihs}\ \emph {et~al.}(1998)\citenamefont {Weihs},
  \citenamefont {Jennewein}, \citenamefont {Simon}, \citenamefont
  {Weinfurter},\ and\ \citenamefont {Zeilinger}}]{weihs_violation_1998}%
  \BibitemOpen
  \bibfield  {author} {\bibinfo {author} {\bibfnamefont {G.}~\bibnamefont
  {Weihs}}, \bibinfo {author} {\bibfnamefont {T.}~\bibnamefont {Jennewein}},
  \bibinfo {author} {\bibfnamefont {C.}~\bibnamefont {Simon}}, \bibinfo
  {author} {\bibfnamefont {H.}~\bibnamefont {Weinfurter}},\ and\ \bibinfo
  {author} {\bibfnamefont {A.}~\bibnamefont {Zeilinger}},\ }\bibfield  {title}
  {\bibinfo {title} {Violation of bell's inequality under strict einstein
  locality conditions},\ }\href {https://doi.org/10.1103/PhysRevLett.81.5039}
  {\bibfield  {journal} {\bibinfo  {journal} {Phys. Rev. Lett.}\ }\textbf
  {\bibinfo {volume} {81}},\ \bibinfo {pages} {5039} (\bibinfo {year}
  {1998})}\BibitemShut {NoStop}%
\bibitem [{\citenamefont {Jennewein}\ \emph {et~al.}(2000)\citenamefont
  {Jennewein}, \citenamefont {Simon}, \citenamefont {Weihs}, \citenamefont
  {Weinfurter},\ and\ \citenamefont {Zeilinger}}]{jennewein_quantum_2000}%
  \BibitemOpen
  \bibfield  {author} {\bibinfo {author} {\bibfnamefont {T.}~\bibnamefont
  {Jennewein}}, \bibinfo {author} {\bibfnamefont {C.}~\bibnamefont {Simon}},
  \bibinfo {author} {\bibfnamefont {G.}~\bibnamefont {Weihs}}, \bibinfo
  {author} {\bibfnamefont {H.}~\bibnamefont {Weinfurter}},\ and\ \bibinfo
  {author} {\bibfnamefont {A.}~\bibnamefont {Zeilinger}},\ }\bibfield  {title}
  {\bibinfo {title} {Quantum {Cryptography} with {Entangled} {Photons}},\
  }\href {https://doi.org/10.1103/PhysRevLett.84.4729} {\bibfield  {journal}
  {\bibinfo  {journal} {Phys. Rev. Lett.}\ }\textbf {\bibinfo {volume} {84}},\
  \bibinfo {pages} {4729} (\bibinfo {year} {2000})}\BibitemShut {NoStop}%
\bibitem [{\citenamefont {Uppu}\ \emph {et~al.}(2021)\citenamefont {Uppu},
  \citenamefont {Midolo}, \citenamefont {Zhou}, \citenamefont {Carolan},\ and\
  \citenamefont {Lodahl}}]{uppu_quantum-dot-based_2021}%
  \BibitemOpen
  \bibfield  {author} {\bibinfo {author} {\bibfnamefont {R.}~\bibnamefont
  {Uppu}}, \bibinfo {author} {\bibfnamefont {L.}~\bibnamefont {Midolo}},
  \bibinfo {author} {\bibfnamefont {X.}~\bibnamefont {Zhou}}, \bibinfo {author}
  {\bibfnamefont {J.}~\bibnamefont {Carolan}},\ and\ \bibinfo {author}
  {\bibfnamefont {P.}~\bibnamefont {Lodahl}},\ }\bibfield  {title} {\bibinfo
  {title} {Quantum-dot-based deterministic photon–emitter interfaces for
  scalable photonic quantum technology},\ }\href
  {https://doi.org/10.1038/s41565-021-00965-6} {\bibfield  {journal} {\bibinfo
  {journal} {Nat. Nanotechnol.}\ }\textbf {\bibinfo {volume} {16}},\ \bibinfo
  {pages} {1308} (\bibinfo {year} {2021})}\BibitemShut {NoStop}%
\bibitem [{\citenamefont {Zhai}\ \emph {et~al.}(2022)\citenamefont {Zhai},
  \citenamefont {Nguyen}, \citenamefont {Spinnler}, \citenamefont {Ritzmann},
  \citenamefont {Löbl}, \citenamefont {Wieck}, \citenamefont {Ludwig},
  \citenamefont {Javadi},\ and\ \citenamefont {Warburton}}]{zhai_quantum_2022}%
  \BibitemOpen
  \bibfield  {author} {\bibinfo {author} {\bibfnamefont {L.}~\bibnamefont
  {Zhai}}, \bibinfo {author} {\bibfnamefont {G.~N.}\ \bibnamefont {Nguyen}},
  \bibinfo {author} {\bibfnamefont {C.}~\bibnamefont {Spinnler}}, \bibinfo
  {author} {\bibfnamefont {J.}~\bibnamefont {Ritzmann}}, \bibinfo {author}
  {\bibfnamefont {M.~C.}\ \bibnamefont {Löbl}}, \bibinfo {author}
  {\bibfnamefont {A.~D.}\ \bibnamefont {Wieck}}, \bibinfo {author}
  {\bibfnamefont {A.}~\bibnamefont {Ludwig}}, \bibinfo {author} {\bibfnamefont
  {A.}~\bibnamefont {Javadi}},\ and\ \bibinfo {author} {\bibfnamefont {R.~J.}\
  \bibnamefont {Warburton}},\ }\bibfield  {title} {\bibinfo {title} {Quantum
  interference of identical photons from remote {GaAs} quantum dots},\ }\href
  {https://doi.org/10.1038/s41565-022-01131-2} {\bibfield  {journal} {\bibinfo
  {journal} {Nat. Nanotechnol.}\ }\textbf {\bibinfo {volume} {17}},\ \bibinfo
  {pages} {829} (\bibinfo {year} {2022})}\BibitemShut {NoStop}%
\bibitem [{\citenamefont {Tomm}\ \emph {et~al.}(2021)\citenamefont {Tomm},
  \citenamefont {Javadi}, \citenamefont {Antoniadis}, \citenamefont {Najer},
  \citenamefont {Löbl}, \citenamefont {Korsch}, \citenamefont {Schott},
  \citenamefont {Valentin}, \citenamefont {Wieck}, \citenamefont {Ludwig},\
  and\ \citenamefont {Warburton}}]{tomm_bright_2021}%
  \BibitemOpen
  \bibfield  {author} {\bibinfo {author} {\bibfnamefont {N.}~\bibnamefont
  {Tomm}}, \bibinfo {author} {\bibfnamefont {A.}~\bibnamefont {Javadi}},
  \bibinfo {author} {\bibfnamefont {N.~O.}\ \bibnamefont {Antoniadis}},
  \bibinfo {author} {\bibfnamefont {D.}~\bibnamefont {Najer}}, \bibinfo
  {author} {\bibfnamefont {M.~C.}\ \bibnamefont {Löbl}}, \bibinfo {author}
  {\bibfnamefont {A.~R.}\ \bibnamefont {Korsch}}, \bibinfo {author}
  {\bibfnamefont {R.}~\bibnamefont {Schott}}, \bibinfo {author} {\bibfnamefont
  {S.~R.}\ \bibnamefont {Valentin}}, \bibinfo {author} {\bibfnamefont {A.~D.}\
  \bibnamefont {Wieck}}, \bibinfo {author} {\bibfnamefont {A.}~\bibnamefont
  {Ludwig}},\ and\ \bibinfo {author} {\bibfnamefont {R.~J.}\ \bibnamefont
  {Warburton}},\ }\bibfield  {title} {\bibinfo {title} {A bright and fast
  source of coherent single photons},\ }\href
  {https://doi.org/10.1038/s41565-020-00831-x} {\bibfield  {journal} {\bibinfo
  {journal} {Nat. Nanotechnol.}\ }\textbf {\bibinfo {volume} {16}},\ \bibinfo
  {pages} {399} (\bibinfo {year} {2021})}\BibitemShut {NoStop}%
\bibitem [{\citenamefont {Liu}\ \emph {et~al.}(2019)\citenamefont {Liu},
  \citenamefont {Su}, \citenamefont {Wei}, \citenamefont {Yao}, \citenamefont
  {Silva}, \citenamefont {Yu}, \citenamefont {Iles-Smith}, \citenamefont
  {Srinivasan}, \citenamefont {Rastelli}, \citenamefont {Li},\ and\
  \citenamefont {Wang}}]{liu_solid-state_2019}%
  \BibitemOpen
  \bibfield  {author} {\bibinfo {author} {\bibfnamefont {J.}~\bibnamefont
  {Liu}}, \bibinfo {author} {\bibfnamefont {R.}~\bibnamefont {Su}}, \bibinfo
  {author} {\bibfnamefont {Y.}~\bibnamefont {Wei}}, \bibinfo {author}
  {\bibfnamefont {B.}~\bibnamefont {Yao}}, \bibinfo {author} {\bibfnamefont
  {S.~F. C.~d.}\ \bibnamefont {Silva}}, \bibinfo {author} {\bibfnamefont
  {Y.}~\bibnamefont {Yu}}, \bibinfo {author} {\bibfnamefont {J.}~\bibnamefont
  {Iles-Smith}}, \bibinfo {author} {\bibfnamefont {K.}~\bibnamefont
  {Srinivasan}}, \bibinfo {author} {\bibfnamefont {A.}~\bibnamefont
  {Rastelli}}, \bibinfo {author} {\bibfnamefont {J.}~\bibnamefont {Li}},\ and\
  \bibinfo {author} {\bibfnamefont {X.}~\bibnamefont {Wang}},\ }\bibfield
  {title} {\bibinfo {title} {A solid-state source of strongly entangled photon
  pairs with high brightness and indistinguishability},\ }\href
  {https://doi.org/10.1038/s41565-019-0435-9} {\bibfield  {journal} {\bibinfo
  {journal} {Nat. Nanotechnol.}\ }\textbf {\bibinfo {volume} {14}},\ \bibinfo
  {pages} {586} (\bibinfo {year} {2019})}\BibitemShut {NoStop}%
\bibitem [{\citenamefont {Vajner}\ \emph {et~al.}(2022)\citenamefont {Vajner},
  \citenamefont {Rickert}, \citenamefont {Gao}, \citenamefont {Kaymazlar},\
  and\ \citenamefont {Heindel}}]{vajner_quantum_2022}%
  \BibitemOpen
  \bibfield  {author} {\bibinfo {author} {\bibfnamefont {D.~A.}\ \bibnamefont
  {Vajner}}, \bibinfo {author} {\bibfnamefont {L.}~\bibnamefont {Rickert}},
  \bibinfo {author} {\bibfnamefont {T.}~\bibnamefont {Gao}}, \bibinfo {author}
  {\bibfnamefont {K.}~\bibnamefont {Kaymazlar}},\ and\ \bibinfo {author}
  {\bibfnamefont {T.}~\bibnamefont {Heindel}},\ }\bibfield  {title} {\bibinfo
  {title} {Quantum {Communication} {Using} {Semiconductor} {Quantum} {Dots}},\
  }\href {https://doi.org/10.1002/qute.202100116} {\bibfield  {journal}
  {\bibinfo  {journal} {Adv. Quantum Technol.}\ }\textbf {\bibinfo {volume}
  {5}},\ \bibinfo {pages} {2100116} (\bibinfo {year} {2022})}\BibitemShut
  {NoStop}%
\bibitem [{\citenamefont {Senellart}\ \emph {et~al.}(2017)\citenamefont
  {Senellart}, \citenamefont {Solomon},\ and\ \citenamefont
  {White}}]{senellart_high-performance_2017}%
  \BibitemOpen
  \bibfield  {author} {\bibinfo {author} {\bibfnamefont {P.}~\bibnamefont
  {Senellart}}, \bibinfo {author} {\bibfnamefont {G.}~\bibnamefont {Solomon}},\
  and\ \bibinfo {author} {\bibfnamefont {A.}~\bibnamefont {White}},\ }\bibfield
   {title} {\bibinfo {title} {High-performance semiconductor quantum-dot
  single-photon sources},\ }\href {https://doi.org/10.1038/nnano.2017.218}
  {\bibfield  {journal} {\bibinfo  {journal} {Nat. Nanotechnol.}\ }\textbf
  {\bibinfo {volume} {12}},\ \bibinfo {pages} {1026} (\bibinfo {year}
  {2017})}\BibitemShut {NoStop}%
\bibitem [{\citenamefont {Schimpf}\ \emph
  {et~al.}(2021{\natexlab{b}})\citenamefont {Schimpf}, \citenamefont {Reindl},
  \citenamefont {Huber}, \citenamefont {Lehner}, \citenamefont {Covre
  Da~Silva}, \citenamefont {Manna}, \citenamefont {Vyvlecka}, \citenamefont
  {Walther},\ and\ \citenamefont {Rastelli}}]{schimpf_quantum_2021}%
  \BibitemOpen
  \bibfield  {author} {\bibinfo {author} {\bibfnamefont {C.}~\bibnamefont
  {Schimpf}}, \bibinfo {author} {\bibfnamefont {M.}~\bibnamefont {Reindl}},
  \bibinfo {author} {\bibfnamefont {D.}~\bibnamefont {Huber}}, \bibinfo
  {author} {\bibfnamefont {B.}~\bibnamefont {Lehner}}, \bibinfo {author}
  {\bibfnamefont {S.~F.}\ \bibnamefont {Covre Da~Silva}}, \bibinfo {author}
  {\bibfnamefont {S.}~\bibnamefont {Manna}}, \bibinfo {author} {\bibfnamefont
  {M.}~\bibnamefont {Vyvlecka}}, \bibinfo {author} {\bibfnamefont
  {P.}~\bibnamefont {Walther}},\ and\ \bibinfo {author} {\bibfnamefont
  {A.}~\bibnamefont {Rastelli}},\ }\bibfield  {title} {\bibinfo {title}
  {Quantum cryptography with highly entangled photons from semiconductor
  quantum dots},\ }\href {https://doi.org/10.1126/sciadv.abe8905} {\bibfield
  {journal} {\bibinfo  {journal} {Sci. Adv.}\ }\textbf {\bibinfo {volume}
  {7}},\ \bibinfo {pages} {eabe8905} (\bibinfo {year}
  {2021}{\natexlab{b}})}\BibitemShut {NoStop}%
\bibitem [{\citenamefont {Huber}\ \emph {et~al.}(2016)\citenamefont {Huber},
  \citenamefont {Ostermann}, \citenamefont {Prilmüller}, \citenamefont
  {Solomon}, \citenamefont {Ritsch}, \citenamefont {Weihs},\ and\ \citenamefont
  {Predojević}}]{huber_coherence_2016}%
  \BibitemOpen
  \bibfield  {author} {\bibinfo {author} {\bibfnamefont {T.}~\bibnamefont
  {Huber}}, \bibinfo {author} {\bibfnamefont {L.}~\bibnamefont {Ostermann}},
  \bibinfo {author} {\bibfnamefont {M.}~\bibnamefont {Prilmüller}}, \bibinfo
  {author} {\bibfnamefont {G.~S.}\ \bibnamefont {Solomon}}, \bibinfo {author}
  {\bibfnamefont {H.}~\bibnamefont {Ritsch}}, \bibinfo {author} {\bibfnamefont
  {G.}~\bibnamefont {Weihs}},\ and\ \bibinfo {author} {\bibfnamefont
  {A.}~\bibnamefont {Predojević}},\ }\bibfield  {title} {\bibinfo {title}
  {Coherence and degree of time-bin entanglement from quantum dots},\ }\href
  {https://doi.org/10.1103/PhysRevB.93.201301} {\bibfield  {journal} {\bibinfo
  {journal} {Phys. Rev. B}\ }\textbf {\bibinfo {volume} {93}},\ \bibinfo
  {pages} {201301} (\bibinfo {year} {2016})}\BibitemShut {NoStop}%
\bibitem [{\citenamefont {Hanschke}\ \emph {et~al.}(2018)\citenamefont
  {Hanschke}, \citenamefont {Fischer}, \citenamefont {Appel}, \citenamefont
  {Lukin}, \citenamefont {Wierzbowski}, \citenamefont {Sun}, \citenamefont
  {Trivedi}, \citenamefont {Vuckovic}, \citenamefont {Finley},\ and\
  \citenamefont {Muller}}]{hanschke_quantum_2018}%
  \BibitemOpen
  \bibfield  {author} {\bibinfo {author} {\bibfnamefont {L.}~\bibnamefont
  {Hanschke}}, \bibinfo {author} {\bibfnamefont {K.~A.}\ \bibnamefont
  {Fischer}}, \bibinfo {author} {\bibfnamefont {S.}~\bibnamefont {Appel}},
  \bibinfo {author} {\bibfnamefont {D.}~\bibnamefont {Lukin}}, \bibinfo
  {author} {\bibfnamefont {J.}~\bibnamefont {Wierzbowski}}, \bibinfo {author}
  {\bibfnamefont {S.}~\bibnamefont {Sun}}, \bibinfo {author} {\bibfnamefont
  {R.}~\bibnamefont {Trivedi}}, \bibinfo {author} {\bibfnamefont
  {J.}~\bibnamefont {Vuckovic}}, \bibinfo {author} {\bibfnamefont {J.~J.}\
  \bibnamefont {Finley}},\ and\ \bibinfo {author} {\bibfnamefont
  {K.}~\bibnamefont {Muller}},\ }\bibfield  {title} {\bibinfo {title} {Quantum
  dot single-photon sources with ultra-low multi-photon probability},\ }\href
  {https://doi.org/10.1038/s41534-018-0092-0} {\bibfield  {journal} {\bibinfo
  {journal} {npj Quantum Inf.}\ }\textbf {\bibinfo {volume} {4}},\ \bibinfo
  {pages} {43} (\bibinfo {year} {2018})}\BibitemShut {NoStop}%
\bibitem [{\citenamefont {Ding}\ \emph {et~al.}(2016)\citenamefont {Ding},
  \citenamefont {He}, \citenamefont {Duan}, \citenamefont {Gregersen},
  \citenamefont {Chen}, \citenamefont {Unsleber}, \citenamefont {Maier},
  \citenamefont {Schneider}, \citenamefont {Kamp}, \citenamefont {Höfling},
  \citenamefont {Lu},\ and\ \citenamefont {Pan}}]{ding_-demand_2016}%
  \BibitemOpen
  \bibfield  {author} {\bibinfo {author} {\bibfnamefont {X.}~\bibnamefont
  {Ding}}, \bibinfo {author} {\bibfnamefont {Y.}~\bibnamefont {He}}, \bibinfo
  {author} {\bibfnamefont {Z.-C.}\ \bibnamefont {Duan}}, \bibinfo {author}
  {\bibfnamefont {N.}~\bibnamefont {Gregersen}}, \bibinfo {author}
  {\bibfnamefont {M.-C.}\ \bibnamefont {Chen}}, \bibinfo {author}
  {\bibfnamefont {S.}~\bibnamefont {Unsleber}}, \bibinfo {author}
  {\bibfnamefont {S.}~\bibnamefont {Maier}}, \bibinfo {author} {\bibfnamefont
  {C.}~\bibnamefont {Schneider}}, \bibinfo {author} {\bibfnamefont
  {M.}~\bibnamefont {Kamp}}, \bibinfo {author} {\bibfnamefont {S.}~\bibnamefont
  {Höfling}}, \bibinfo {author} {\bibfnamefont {C.-Y.}\ \bibnamefont {Lu}},\
  and\ \bibinfo {author} {\bibfnamefont {J.-W.}\ \bibnamefont {Pan}},\
  }\bibfield  {title} {\bibinfo {title} {On-{Demand} {Single} {Photons} with
  {High} {Extraction} {Efficiency} and {Near}-{Unity} {Indistinguishability}
  from a {Resonantly} {Driven} {Quantum} {Dot} in a {Micropillar}},\ }\href
  {https://doi.org/10.1103/PhysRevLett.116.020401} {\bibfield  {journal}
  {\bibinfo  {journal} {Phys. Rev. Lett.}\ }\textbf {\bibinfo {volume} {116}},\
  \bibinfo {pages} {020401} (\bibinfo {year} {2016})}\BibitemShut {NoStop}%
\bibitem [{\citenamefont {Zhao}\ \emph {et~al.}(2020)\citenamefont {Zhao},
  \citenamefont {Chen}, \citenamefont {Yu}, \citenamefont {Li}, \citenamefont
  {Davanco},\ and\ \citenamefont {Liu}}]{zhao_advanced_2020}%
  \BibitemOpen
  \bibfield  {author} {\bibinfo {author} {\bibfnamefont {T.~M.}\ \bibnamefont
  {Zhao}}, \bibinfo {author} {\bibfnamefont {Y.}~\bibnamefont {Chen}}, \bibinfo
  {author} {\bibfnamefont {Y.}~\bibnamefont {Yu}}, \bibinfo {author}
  {\bibfnamefont {Q.}~\bibnamefont {Li}}, \bibinfo {author} {\bibfnamefont
  {M.}~\bibnamefont {Davanco}},\ and\ \bibinfo {author} {\bibfnamefont
  {J.}~\bibnamefont {Liu}},\ }\bibfield  {title} {\bibinfo {title} {Advanced
  {Technologies} for {Quantum} {Photonic} {Devices} {Based} on {Epitaxial}
  {Quantum} {Dots}},\ }\href {https://doi.org/10.1002/qute.201900034}
  {\bibfield  {journal} {\bibinfo  {journal} {Adv. Quantum Technol.}\ }\textbf
  {\bibinfo {volume} {3}},\ \bibinfo {pages} {1900034} (\bibinfo {year}
  {2020})}\BibitemShut {NoStop}%
\bibitem [{\citenamefont {Grim}\ \emph {et~al.}(2019)\citenamefont {Grim},
  \citenamefont {Bracker}, \citenamefont {Zalalutdinov}, \citenamefont
  {Carter}, \citenamefont {Kozen}, \citenamefont {Kim}, \citenamefont {Kim},
  \citenamefont {Mlack}, \citenamefont {Yakes}, \citenamefont {Lee},\ and\
  \citenamefont {Gammon}}]{grim_scalable_2019}%
  \BibitemOpen
  \bibfield  {author} {\bibinfo {author} {\bibfnamefont {J.~Q.}\ \bibnamefont
  {Grim}}, \bibinfo {author} {\bibfnamefont {A.~S.}\ \bibnamefont {Bracker}},
  \bibinfo {author} {\bibfnamefont {M.}~\bibnamefont {Zalalutdinov}}, \bibinfo
  {author} {\bibfnamefont {S.~G.}\ \bibnamefont {Carter}}, \bibinfo {author}
  {\bibfnamefont {A.~C.}\ \bibnamefont {Kozen}}, \bibinfo {author}
  {\bibfnamefont {M.}~\bibnamefont {Kim}}, \bibinfo {author} {\bibfnamefont
  {C.~S.}\ \bibnamefont {Kim}}, \bibinfo {author} {\bibfnamefont {J.~T.}\
  \bibnamefont {Mlack}}, \bibinfo {author} {\bibfnamefont {M.}~\bibnamefont
  {Yakes}}, \bibinfo {author} {\bibfnamefont {B.}~\bibnamefont {Lee}},\ and\
  \bibinfo {author} {\bibfnamefont {D.}~\bibnamefont {Gammon}},\ }\bibfield
  {title} {\bibinfo {title} {Scalable in operando strain tuning in nanophotonic
  waveguides enabling three-quantum-dot superradiance},\ }\href
  {https://doi.org/10.1038/s41563-019-0418-0} {\bibfield  {journal} {\bibinfo
  {journal} {Nature Materials}\ }\textbf {\bibinfo {volume} {18}},\ \bibinfo
  {pages} {963} (\bibinfo {year} {2019})}\BibitemShut {NoStop}%
\bibitem [{\citenamefont {Ramsay}(2010)}]{ramsay_review_2010}%
  \BibitemOpen
  \bibfield  {author} {\bibinfo {author} {\bibfnamefont {A.~J.}\ \bibnamefont
  {Ramsay}},\ }\bibfield  {title} {\bibinfo {title} {A review of the coherent
  optical control of the exciton and spin states of semiconductor quantum
  dots},\ }\href {https://doi.org/10.1088/0268-1242/25/10/103001} {\bibfield
  {journal} {\bibinfo  {journal} {Semicond. Sci. Technol.}\ }\textbf {\bibinfo
  {volume} {25}},\ \bibinfo {pages} {103001} (\bibinfo {year}
  {2010})}\BibitemShut {NoStop}%
\bibitem [{\citenamefont {Stufler}\ \emph {et~al.}(2006)\citenamefont
  {Stufler}, \citenamefont {Machnikowski}, \citenamefont {Ester}, \citenamefont
  {Bichler}, \citenamefont {Axt}, \citenamefont {Kuhn},\ and\ \citenamefont
  {Zrenner}}]{stufler_two-photon_2006}%
  \BibitemOpen
  \bibfield  {author} {\bibinfo {author} {\bibfnamefont {S.}~\bibnamefont
  {Stufler}}, \bibinfo {author} {\bibfnamefont {P.}~\bibnamefont
  {Machnikowski}}, \bibinfo {author} {\bibfnamefont {P.}~\bibnamefont {Ester}},
  \bibinfo {author} {\bibfnamefont {M.}~\bibnamefont {Bichler}}, \bibinfo
  {author} {\bibfnamefont {V.~M.}\ \bibnamefont {Axt}}, \bibinfo {author}
  {\bibfnamefont {T.}~\bibnamefont {Kuhn}},\ and\ \bibinfo {author}
  {\bibfnamefont {A.}~\bibnamefont {Zrenner}},\ }\bibfield  {title} {\bibinfo
  {title} {Two-photon {Rabi} oscillations in a single
  $\mathrm{In}_{x}\mathrm{Ga}_{1-x}\mathrm{As}\mathrm{/}\mathrm{Ga}\mathrm{As}$
  quantum dot},\ }\href {https://doi.org/10.1103/PhysRevB.73.125304} {\bibfield
   {journal} {\bibinfo  {journal} {Phys. Rev. B}\ }\textbf {\bibinfo {volume}
  {73}},\ \bibinfo {pages} {125304} (\bibinfo {year} {2006})}\BibitemShut
  {NoStop}%
\bibitem [{\citenamefont {Jayakumar}\ \emph {et~al.}(2013)\citenamefont
  {Jayakumar}, \citenamefont {Predojević}, \citenamefont {Huber},
  \citenamefont {Kauten}, \citenamefont {Solomon},\ and\ \citenamefont
  {Weihs}}]{jayakumar_deterministic_2013}%
  \BibitemOpen
  \bibfield  {author} {\bibinfo {author} {\bibfnamefont {H.}~\bibnamefont
  {Jayakumar}}, \bibinfo {author} {\bibfnamefont {A.}~\bibnamefont
  {Predojević}}, \bibinfo {author} {\bibfnamefont {T.}~\bibnamefont {Huber}},
  \bibinfo {author} {\bibfnamefont {T.}~\bibnamefont {Kauten}}, \bibinfo
  {author} {\bibfnamefont {G.~S.}\ \bibnamefont {Solomon}},\ and\ \bibinfo
  {author} {\bibfnamefont {G.}~\bibnamefont {Weihs}},\ }\bibfield  {title}
  {\bibinfo {title} {Deterministic {Photon} {Pairs} and {Coherent} {Optical}
  {Control} of a {Single} {Quantum} {Dot}},\ }\href
  {https://doi.org/10.1103/PhysRevLett.110.135505} {\bibfield  {journal}
  {\bibinfo  {journal} {Phys. Rev. Lett.}\ }\textbf {\bibinfo {volume} {110}},\
  \bibinfo {pages} {135505} (\bibinfo {year} {2013})}\BibitemShut {NoStop}%
\bibitem [{\citenamefont {Simon}\ \emph {et~al.}(2011)\citenamefont {Simon},
  \citenamefont {Belhadj}, \citenamefont {Chatel}, \citenamefont {Amand},
  \citenamefont {Renucci}, \citenamefont {Lemaitre}, \citenamefont {Krebs},
  \citenamefont {Dalgarno}, \citenamefont {Warburton}, \citenamefont {Marie},\
  and\ \citenamefont {Urbaszek}}]{simon_robust_2011}%
  \BibitemOpen
  \bibfield  {author} {\bibinfo {author} {\bibfnamefont {C.-M.}\ \bibnamefont
  {Simon}}, \bibinfo {author} {\bibfnamefont {T.}~\bibnamefont {Belhadj}},
  \bibinfo {author} {\bibfnamefont {B.}~\bibnamefont {Chatel}}, \bibinfo
  {author} {\bibfnamefont {T.}~\bibnamefont {Amand}}, \bibinfo {author}
  {\bibfnamefont {P.}~\bibnamefont {Renucci}}, \bibinfo {author} {\bibfnamefont
  {A.}~\bibnamefont {Lemaitre}}, \bibinfo {author} {\bibfnamefont
  {O.}~\bibnamefont {Krebs}}, \bibinfo {author} {\bibfnamefont {P.~A.}\
  \bibnamefont {Dalgarno}}, \bibinfo {author} {\bibfnamefont {R.~J.}\
  \bibnamefont {Warburton}}, \bibinfo {author} {\bibfnamefont {X.}~\bibnamefont
  {Marie}},\ and\ \bibinfo {author} {\bibfnamefont {B.}~\bibnamefont
  {Urbaszek}},\ }\bibfield  {title} {\bibinfo {title} {Robust {Quantum} {Dot}
  {Exciton} {Generation} via {Adiabatic} {Passage} with {Frequency}-{Swept}
  {Optical} {Pulses}},\ }\href {https://doi.org/10.1103/PhysRevLett.106.166801}
  {\bibfield  {journal} {\bibinfo  {journal} {Phys. Rev. Lett.}\ }\textbf
  {\bibinfo {volume} {106}},\ \bibinfo {pages} {166801} (\bibinfo {year}
  {2011})}\BibitemShut {NoStop}%
\bibitem [{\citenamefont {Wu}\ \emph {et~al.}(2011)\citenamefont {Wu},
  \citenamefont {Piper}, \citenamefont {Ediger}, \citenamefont {Brereton},
  \citenamefont {Schmidgall}, \citenamefont {Eastham}, \citenamefont {Hugues},
  \citenamefont {Hopkinson},\ and\ \citenamefont
  {Phillips}}]{wu_population_2011}%
  \BibitemOpen
  \bibfield  {author} {\bibinfo {author} {\bibfnamefont {Y.}~\bibnamefont
  {Wu}}, \bibinfo {author} {\bibfnamefont {I.~M.}\ \bibnamefont {Piper}},
  \bibinfo {author} {\bibfnamefont {M.}~\bibnamefont {Ediger}}, \bibinfo
  {author} {\bibfnamefont {P.}~\bibnamefont {Brereton}}, \bibinfo {author}
  {\bibfnamefont {E.~R.}\ \bibnamefont {Schmidgall}}, \bibinfo {author}
  {\bibfnamefont {P.~R.}\ \bibnamefont {Eastham}}, \bibinfo {author}
  {\bibfnamefont {M.}~\bibnamefont {Hugues}}, \bibinfo {author} {\bibfnamefont
  {M.}~\bibnamefont {Hopkinson}},\ and\ \bibinfo {author} {\bibfnamefont
  {R.~T.}\ \bibnamefont {Phillips}},\ }\bibfield  {title} {\bibinfo {title}
  {Population {Inversion} in a {Single} {InGaAs} {Quantum} {Dot} {Using} the
  {Method} of {Adiabatic} {Rapid} {Passage}},\ }\href
  {https://doi.org/10.1103/PhysRevLett.106.067401} {\bibfield  {journal}
  {\bibinfo  {journal} {Phys. Rev. Lett.}\ }\textbf {\bibinfo {volume} {106}},\
  \bibinfo {pages} {067401} (\bibinfo {year} {2011})}\BibitemShut {NoStop}%
\bibitem [{\citenamefont {Debnath}\ \emph {et~al.}(2013)\citenamefont
  {Debnath}, \citenamefont {Meier}, \citenamefont {Chatel},\ and\ \citenamefont
  {Amand}}]{debnath_high-fidelity_2013}%
  \BibitemOpen
  \bibfield  {author} {\bibinfo {author} {\bibfnamefont {A.}~\bibnamefont
  {Debnath}}, \bibinfo {author} {\bibfnamefont {C.}~\bibnamefont {Meier}},
  \bibinfo {author} {\bibfnamefont {B.}~\bibnamefont {Chatel}},\ and\ \bibinfo
  {author} {\bibfnamefont {T.}~\bibnamefont {Amand}},\ }\bibfield  {title}
  {\bibinfo {title} {High-fidelity biexciton generation in quantum dots by
  chirped laser pulses},\ }\href {https://doi.org/10.1103/PhysRevB.88.201305}
  {\bibfield  {journal} {\bibinfo  {journal} {Phys. Rev. B}\ }\textbf {\bibinfo
  {volume} {88}},\ \bibinfo {pages} {201305} (\bibinfo {year}
  {2013})}\BibitemShut {NoStop}%
\bibitem [{\citenamefont {Kaldewey}\ \emph {et~al.}(2017)\citenamefont
  {Kaldewey}, \citenamefont {Lüker}, \citenamefont {Kuhlmann}, \citenamefont
  {Valentin}, \citenamefont {Ludwig}, \citenamefont {Wieck}, \citenamefont
  {Reiter}, \citenamefont {Kuhn},\ and\ \citenamefont
  {Warburton}}]{kaldewey_coherent_2017}%
  \BibitemOpen
  \bibfield  {author} {\bibinfo {author} {\bibfnamefont {T.}~\bibnamefont
  {Kaldewey}}, \bibinfo {author} {\bibfnamefont {S.}~\bibnamefont {Lüker}},
  \bibinfo {author} {\bibfnamefont {A.~V.}\ \bibnamefont {Kuhlmann}}, \bibinfo
  {author} {\bibfnamefont {S.~R.}\ \bibnamefont {Valentin}}, \bibinfo {author}
  {\bibfnamefont {A.}~\bibnamefont {Ludwig}}, \bibinfo {author} {\bibfnamefont
  {A.~D.}\ \bibnamefont {Wieck}}, \bibinfo {author} {\bibfnamefont {D.~E.}\
  \bibnamefont {Reiter}}, \bibinfo {author} {\bibfnamefont {T.}~\bibnamefont
  {Kuhn}},\ and\ \bibinfo {author} {\bibfnamefont {R.~J.}\ \bibnamefont
  {Warburton}},\ }\bibfield  {title} {\bibinfo {title} {Coherent and robust
  high-fidelity generation of a biexciton in a quantum dot by rapid adiabatic
  passage},\ }\href {https://doi.org/10.1103/PhysRevB.95.161302} {\bibfield
  {journal} {\bibinfo  {journal} {Phys. Rev. B}\ }\textbf {\bibinfo {volume}
  {95}},\ \bibinfo {pages} {161302(R)} (\bibinfo {year} {2017})}\BibitemShut
  {NoStop}%
\bibitem [{\citenamefont {Mathew}\ \emph {et~al.}(2014)\citenamefont {Mathew},
  \citenamefont {Dilcher}, \citenamefont {Gamouras}, \citenamefont
  {Ramachandran}, \citenamefont {Yang}, \citenamefont {Freisem}, \citenamefont
  {Deppe},\ and\ \citenamefont {Hall}}]{mathew_subpicosecond_2014}%
  \BibitemOpen
  \bibfield  {author} {\bibinfo {author} {\bibfnamefont {R.}~\bibnamefont
  {Mathew}}, \bibinfo {author} {\bibfnamefont {E.}~\bibnamefont {Dilcher}},
  \bibinfo {author} {\bibfnamefont {A.}~\bibnamefont {Gamouras}}, \bibinfo
  {author} {\bibfnamefont {A.}~\bibnamefont {Ramachandran}}, \bibinfo {author}
  {\bibfnamefont {H.~Y.~S.}\ \bibnamefont {Yang}}, \bibinfo {author}
  {\bibfnamefont {S.}~\bibnamefont {Freisem}}, \bibinfo {author} {\bibfnamefont
  {D.}~\bibnamefont {Deppe}},\ and\ \bibinfo {author} {\bibfnamefont {K.~C.}\
  \bibnamefont {Hall}},\ }\bibfield  {title} {\bibinfo {title} {Subpicosecond
  adiabatic rapid passage on a single semiconductor quantum dot:
  {Phonon}-mediated dephasing in the strong-driving regime},\ }\href
  {https://doi.org/10.1103/PhysRevB.90.035316} {\bibfield  {journal} {\bibinfo
  {journal} {Phys. Rev. B}\ }\textbf {\bibinfo {volume} {90}},\ \bibinfo
  {pages} {035316} (\bibinfo {year} {2014})}\BibitemShut {NoStop}%
\bibitem [{\citenamefont {Creatore}\ \emph {et~al.}(2012)\citenamefont
  {Creatore}, \citenamefont {Brierley}, \citenamefont {Phillips}, \citenamefont
  {Littlewood},\ and\ \citenamefont {Eastham}}]{creatore_creation_2012}%
  \BibitemOpen
  \bibfield  {author} {\bibinfo {author} {\bibfnamefont {C.}~\bibnamefont
  {Creatore}}, \bibinfo {author} {\bibfnamefont {R.~T.}\ \bibnamefont
  {Brierley}}, \bibinfo {author} {\bibfnamefont {R.~T.}\ \bibnamefont
  {Phillips}}, \bibinfo {author} {\bibfnamefont {P.~B.}\ \bibnamefont
  {Littlewood}},\ and\ \bibinfo {author} {\bibfnamefont {P.~R.}\ \bibnamefont
  {Eastham}},\ }\bibfield  {title} {\bibinfo {title} {Creation of entangled
  states in coupled quantum dots via adiabatic rapid passage},\ }\href
  {https://doi.org/10.1103/PhysRevB.86.155442} {\bibfield  {journal} {\bibinfo
  {journal} {Phys. Rev. B}\ }\textbf {\bibinfo {volume} {86}},\ \bibinfo
  {pages} {155442} (\bibinfo {year} {2012})}\BibitemShut {NoStop}%
\bibitem [{\citenamefont {Gamouras}\ \emph {et~al.}(2013)\citenamefont
  {Gamouras}, \citenamefont {Mathew}, \citenamefont {Freisem}, \citenamefont
  {Deppe},\ and\ \citenamefont {Hall}}]{gamouras2013simultaneous}%
  \BibitemOpen
  \bibfield  {author} {\bibinfo {author} {\bibfnamefont {A.}~\bibnamefont
  {Gamouras}}, \bibinfo {author} {\bibfnamefont {R.}~\bibnamefont {Mathew}},
  \bibinfo {author} {\bibfnamefont {S.}~\bibnamefont {Freisem}}, \bibinfo
  {author} {\bibfnamefont {D.~G.}\ \bibnamefont {Deppe}},\ and\ \bibinfo
  {author} {\bibfnamefont {K.~C.}\ \bibnamefont {Hall}},\ }\bibfield  {title}
  {\bibinfo {title} {{Simultaneous deterministic control of distant qubits in
  two semiconductor quantum dots}},\ }\href {https://doi.org/10.1021/nl4018176}
  {\bibfield  {journal} {\bibinfo  {journal} {Nano Lett.}\ }\textbf {\bibinfo
  {volume} {13}},\ \bibinfo {pages} {4666} (\bibinfo {year}
  {2013})}\BibitemShut {NoStop}%
\bibitem [{\citenamefont {Ramachandran}\ \emph {et~al.}(2021)\citenamefont
  {Ramachandran}, \citenamefont {Fraser-Leach}, \citenamefont {O’Neal},
  \citenamefont {Deppe},\ and\ \citenamefont
  {Hall}}]{ramachandran_experimental_2021}%
  \BibitemOpen
  \bibfield  {author} {\bibinfo {author} {\bibfnamefont {A.}~\bibnamefont
  {Ramachandran}}, \bibinfo {author} {\bibfnamefont {J.}~\bibnamefont
  {Fraser-Leach}}, \bibinfo {author} {\bibfnamefont {S.}~\bibnamefont
  {O’Neal}}, \bibinfo {author} {\bibfnamefont {D.~G.}\ \bibnamefont
  {Deppe}},\ and\ \bibinfo {author} {\bibfnamefont {K.~C.}\ \bibnamefont
  {Hall}},\ }\bibfield  {title} {\bibinfo {title} {Experimental quantification
  of the robustness of adiabatic rapid passage for quantum state inversion in
  semiconductor quantum dots},\ }\href {https://doi.org/10.1364/OE.435109}
  {\bibfield  {journal} {\bibinfo  {journal} {Opt. Express}\ }\textbf {\bibinfo
  {volume} {29}},\ \bibinfo {pages} {41766} (\bibinfo {year}
  {2021})}\BibitemShut {NoStop}%
\bibitem [{\citenamefont {Reiter}\ \emph {et~al.}(2012)\citenamefont {Reiter},
  \citenamefont {Lüker}, \citenamefont {Gawarecki}, \citenamefont
  {Grodecka-Grad}, \citenamefont {Machnikowski}, \citenamefont {Axt},\ and\
  \citenamefont {Kuhn}}]{reiter_phonon_2012}%
  \BibitemOpen
  \bibfield  {author} {\bibinfo {author} {\bibfnamefont {D.}~\bibnamefont
  {Reiter}}, \bibinfo {author} {\bibfnamefont {S.}~\bibnamefont {Lüker}},
  \bibinfo {author} {\bibfnamefont {K.}~\bibnamefont {Gawarecki}}, \bibinfo
  {author} {\bibfnamefont {A.}~\bibnamefont {Grodecka-Grad}}, \bibinfo {author}
  {\bibfnamefont {P.}~\bibnamefont {Machnikowski}}, \bibinfo {author}
  {\bibfnamefont {V.}~\bibnamefont {Axt}},\ and\ \bibinfo {author}
  {\bibfnamefont {T.}~\bibnamefont {Kuhn}},\ }\bibfield  {title} {\bibinfo
  {title} {Phonon {Effects} on {Population} {Inversion} in {Quantum} {Dots}:
  {Resonant}, {Detuned} and {Frequency}-{Swept} {Excitations}},\ }\href
  {https://doi.org/10.12693/APhysPolA.122.1065} {\bibfield  {journal} {\bibinfo
   {journal} {Acta Phys. Pol.}\ }\textbf {\bibinfo {volume} {122}},\ \bibinfo
  {pages} {1065} (\bibinfo {year} {2012})}\BibitemShut {NoStop}%
\bibitem [{\citenamefont {Ardelt}\ \emph {et~al.}(2014)\citenamefont {Ardelt},
  \citenamefont {Hanschke}, \citenamefont {Fischer}, \citenamefont {Müller},
  \citenamefont {Kleinkauf}, \citenamefont {Koller}, \citenamefont {Bechtold},
  \citenamefont {Simmet}, \citenamefont {Wierzbowski}, \citenamefont {Riedl},
  \citenamefont {Abstreiter},\ and\ \citenamefont
  {Finley}}]{ardelt_dissipative_2014}%
  \BibitemOpen
  \bibfield  {author} {\bibinfo {author} {\bibfnamefont {P.-L.}\ \bibnamefont
  {Ardelt}}, \bibinfo {author} {\bibfnamefont {L.}~\bibnamefont {Hanschke}},
  \bibinfo {author} {\bibfnamefont {K.~A.}\ \bibnamefont {Fischer}}, \bibinfo
  {author} {\bibfnamefont {K.}~\bibnamefont {Müller}}, \bibinfo {author}
  {\bibfnamefont {A.}~\bibnamefont {Kleinkauf}}, \bibinfo {author}
  {\bibfnamefont {M.}~\bibnamefont {Koller}}, \bibinfo {author} {\bibfnamefont
  {A.}~\bibnamefont {Bechtold}}, \bibinfo {author} {\bibfnamefont
  {T.}~\bibnamefont {Simmet}}, \bibinfo {author} {\bibfnamefont
  {J.}~\bibnamefont {Wierzbowski}}, \bibinfo {author} {\bibfnamefont
  {H.}~\bibnamefont {Riedl}}, \bibinfo {author} {\bibfnamefont
  {G.}~\bibnamefont {Abstreiter}},\ and\ \bibinfo {author} {\bibfnamefont
  {J.~J.}\ \bibnamefont {Finley}},\ }\bibfield  {title} {\bibinfo {title}
  {Dissipative preparation of the exciton and biexciton in self-assembled
  quantum dots on picosecond time scales},\ }\href
  {https://doi.org/10.1103/PhysRevB.90.241404} {\bibfield  {journal} {\bibinfo
  {journal} {Phys. Rev. B}\ }\textbf {\bibinfo {volume} {90}},\ \bibinfo
  {pages} {241404} (\bibinfo {year} {2014})}\BibitemShut {NoStop}%
\bibitem [{\citenamefont {Bounouar}\ \emph {et~al.}(2015)\citenamefont
  {Bounouar}, \citenamefont {Müller}, \citenamefont {Barth}, \citenamefont
  {Glässl}, \citenamefont {Axt},\ and\ \citenamefont
  {Michler}}]{bounouar_phonon-assisted_2015}%
  \BibitemOpen
  \bibfield  {author} {\bibinfo {author} {\bibfnamefont {S.}~\bibnamefont
  {Bounouar}}, \bibinfo {author} {\bibfnamefont {M.}~\bibnamefont {Müller}},
  \bibinfo {author} {\bibfnamefont {A.~M.}\ \bibnamefont {Barth}}, \bibinfo
  {author} {\bibfnamefont {M.}~\bibnamefont {Glässl}}, \bibinfo {author}
  {\bibfnamefont {V.~M.}\ \bibnamefont {Axt}},\ and\ \bibinfo {author}
  {\bibfnamefont {P.}~\bibnamefont {Michler}},\ }\bibfield  {title} {\bibinfo
  {title} {Phonon-assisted robust and deterministic two-photon biexciton
  preparation in a quantum dot},\ }\href
  {https://doi.org/10.1103/PhysRevB.91.161302} {\bibfield  {journal} {\bibinfo
  {journal} {Phys. Rev. B}\ }\textbf {\bibinfo {volume} {91}},\ \bibinfo
  {pages} {161302} (\bibinfo {year} {2015})}\BibitemShut {NoStop}%
\bibitem [{\citenamefont {Quilter}\ \emph {et~al.}(2015)\citenamefont
  {Quilter}, \citenamefont {Brash}, \citenamefont {Liu}, \citenamefont
  {Glässl}, \citenamefont {Barth}, \citenamefont {Axt}, \citenamefont
  {Ramsay}, \citenamefont {Skolnick},\ and\ \citenamefont
  {Fox}}]{quilter_phonon-assisted_2015}%
  \BibitemOpen
  \bibfield  {author} {\bibinfo {author} {\bibfnamefont {J.}~\bibnamefont
  {Quilter}}, \bibinfo {author} {\bibfnamefont {A.}~\bibnamefont {Brash}},
  \bibinfo {author} {\bibfnamefont {F.}~\bibnamefont {Liu}}, \bibinfo {author}
  {\bibfnamefont {M.}~\bibnamefont {Glässl}}, \bibinfo {author} {\bibfnamefont
  {A.}~\bibnamefont {Barth}}, \bibinfo {author} {\bibfnamefont
  {V.}~\bibnamefont {Axt}}, \bibinfo {author} {\bibfnamefont {A.}~\bibnamefont
  {Ramsay}}, \bibinfo {author} {\bibfnamefont {M.}~\bibnamefont {Skolnick}},\
  and\ \bibinfo {author} {\bibfnamefont {A.}~\bibnamefont {Fox}},\ }\bibfield
  {title} {\bibinfo {title} {Phonon-{Assisted} {Population} {Inversion} of a
  {Single} {InGaAs} / {GaAs} {Quantum} {Dot} by {Pulsed} {Laser}
  {Excitation}},\ }\href {https://doi.org/10.1103/PhysRevLett.114.137401}
  {\bibfield  {journal} {\bibinfo  {journal} {Phys. Rev. Lett.}\ }\textbf
  {\bibinfo {volume} {114}},\ \bibinfo {pages} {137401} (\bibinfo {year}
  {2015})}\BibitemShut {NoStop}%
\bibitem [{\citenamefont {Reindl}\ \emph {et~al.}(2017)\citenamefont {Reindl},
  \citenamefont {Jöns}, \citenamefont {Huber}, \citenamefont {Schimpf},
  \citenamefont {Huo}, \citenamefont {Zwiller}, \citenamefont {Rastelli},\ and\
  \citenamefont {Trotta}}]{reindl_phonon-assisted_2017}%
  \BibitemOpen
  \bibfield  {author} {\bibinfo {author} {\bibfnamefont {M.}~\bibnamefont
  {Reindl}}, \bibinfo {author} {\bibfnamefont {K.~D.}\ \bibnamefont {Jöns}},
  \bibinfo {author} {\bibfnamefont {D.}~\bibnamefont {Huber}}, \bibinfo
  {author} {\bibfnamefont {C.}~\bibnamefont {Schimpf}}, \bibinfo {author}
  {\bibfnamefont {Y.}~\bibnamefont {Huo}}, \bibinfo {author} {\bibfnamefont
  {V.}~\bibnamefont {Zwiller}}, \bibinfo {author} {\bibfnamefont
  {A.}~\bibnamefont {Rastelli}},\ and\ \bibinfo {author} {\bibfnamefont
  {R.}~\bibnamefont {Trotta}},\ }\bibfield  {title} {\bibinfo {title}
  {Phonon-assisted two-photon interference from remote quantum emitters},\
  }\href {https://doi.org/10.1021/acs.nanolett.7b00777} {\bibfield  {journal}
  {\bibinfo  {journal} {Nano Lett.}\ }\textbf {\bibinfo {volume} {17}},\
  \bibinfo {pages} {4090} (\bibinfo {year} {2017})}\BibitemShut {NoStop}%
\bibitem [{\citenamefont {Glässl}\ \emph
  {et~al.}(2013{\natexlab{a}})\citenamefont {Glässl}, \citenamefont {Barth},\
  and\ \citenamefont {Axt}}]{glassl_proposed_2013}%
  \BibitemOpen
  \bibfield  {author} {\bibinfo {author} {\bibfnamefont {M.}~\bibnamefont
  {Glässl}}, \bibinfo {author} {\bibfnamefont {A.~M.}\ \bibnamefont {Barth}},\
  and\ \bibinfo {author} {\bibfnamefont {V.~M.}\ \bibnamefont {Axt}},\
  }\bibfield  {title} {\bibinfo {title} {Proposed {Robust} and
  {High}-{Fidelity} {Preparation} of {Excitons} and {Biexcitons} in
  {Semiconductor} {Quantum} {Dots} {Making} {Active} {Use} of {Phonons}},\
  }\href {https://doi.org/10.1103/PhysRevLett.110.147401} {\bibfield  {journal}
  {\bibinfo  {journal} {Phys. Rev. Lett.}\ }\textbf {\bibinfo {volume} {110}},\
  \bibinfo {pages} {147401} (\bibinfo {year} {2013}{\natexlab{a}})}\BibitemShut
  {NoStop}%
\bibitem [{\citenamefont {Thomas}\ \emph {et~al.}(2021)\citenamefont {Thomas},
  \citenamefont {Billard}, \citenamefont {Coste}, \citenamefont {Wein},
  \citenamefont {{Priya}}, \citenamefont {Ollivier}, \citenamefont {Krebs},
  \citenamefont {Tazaïrt}, \citenamefont {Harouri}, \citenamefont {Lemaitre},
  \citenamefont {Sagnes}, \citenamefont {Anton}, \citenamefont {Lanco},
  \citenamefont {Somaschi}, \citenamefont {Loredo},\ and\ \citenamefont
  {Senellart}}]{thomas_bright_2021}%
  \BibitemOpen
  \bibfield  {author} {\bibinfo {author} {\bibfnamefont {S.~E.}\ \bibnamefont
  {Thomas}}, \bibinfo {author} {\bibfnamefont {M.}~\bibnamefont {Billard}},
  \bibinfo {author} {\bibfnamefont {N.}~\bibnamefont {Coste}}, \bibinfo
  {author} {\bibfnamefont {S.~C.}\ \bibnamefont {Wein}}, \bibinfo {author}
  {\bibnamefont {{Priya}}}, \bibinfo {author} {\bibfnamefont {H.}~\bibnamefont
  {Ollivier}}, \bibinfo {author} {\bibfnamefont {O.}~\bibnamefont {Krebs}},
  \bibinfo {author} {\bibfnamefont {L.}~\bibnamefont {Tazaïrt}}, \bibinfo
  {author} {\bibfnamefont {A.}~\bibnamefont {Harouri}}, \bibinfo {author}
  {\bibfnamefont {A.}~\bibnamefont {Lemaitre}}, \bibinfo {author}
  {\bibfnamefont {I.}~\bibnamefont {Sagnes}}, \bibinfo {author} {\bibfnamefont
  {C.}~\bibnamefont {Anton}}, \bibinfo {author} {\bibfnamefont
  {L.}~\bibnamefont {Lanco}}, \bibinfo {author} {\bibfnamefont
  {N.}~\bibnamefont {Somaschi}}, \bibinfo {author} {\bibfnamefont {J.~C.}\
  \bibnamefont {Loredo}},\ and\ \bibinfo {author} {\bibfnamefont
  {P.}~\bibnamefont {Senellart}},\ }\bibfield  {title} {\bibinfo {title}
  {Bright {Polarized} {Single}-{Photon} {Source} {Based} on a {Linear}
  {Dipole}},\ }\href {https://doi.org/10.1103/PhysRevLett.126.233601}
  {\bibfield  {journal} {\bibinfo  {journal} {Phys. Rev. Lett.}\ }\textbf
  {\bibinfo {volume} {126}},\ \bibinfo {pages} {233601} (\bibinfo {year}
  {2021})}\BibitemShut {NoStop}%
\bibitem [{\citenamefont {Backus}\ \emph {et~al.}(1998)\citenamefont {Backus},
  \citenamefont {Durfee}, \citenamefont {Murnane},\ and\ \citenamefont
  {Kapteyn}}]{backus_high_1998}%
  \BibitemOpen
  \bibfield  {author} {\bibinfo {author} {\bibfnamefont {S.}~\bibnamefont
  {Backus}}, \bibinfo {author} {\bibfnamefont {C.~G.}\ \bibnamefont {Durfee}},
  \bibinfo {author} {\bibfnamefont {M.~M.}\ \bibnamefont {Murnane}},\ and\
  \bibinfo {author} {\bibfnamefont {H.~C.}\ \bibnamefont {Kapteyn}},\
  }\bibfield  {title} {\bibinfo {title} {High power ultrafast lasers},\ }\href
  {https://doi.org/10.1063/1.1148795} {\bibfield  {journal} {\bibinfo
  {journal} {Rev. Sci. Instrum.}\ }\textbf {\bibinfo {volume} {69}},\ \bibinfo
  {pages} {1207} (\bibinfo {year} {1998})}\BibitemShut {NoStop}%
\bibitem [{\citenamefont {Martinez}(1987)}]{martinez_3000_1987}%
  \BibitemOpen
  \bibfield  {author} {\bibinfo {author} {\bibfnamefont {O.}~\bibnamefont
  {Martinez}},\ }\bibfield  {title} {\bibinfo {title} {3000 times grating
  compressor with positive group velocity dispersion: {Application} to fiber
  compensation in 1.3-1.6 µm region},\ }\href
  {https://doi.org/10.1109/JQE.1987.1073201} {\bibfield  {journal} {\bibinfo
  {journal} {IEEE J. Quant. Elect.}\ }\textbf {\bibinfo {volume} {23}},\
  \bibinfo {pages} {59} (\bibinfo {year} {1987})}\BibitemShut {NoStop}%
\bibitem [{\citenamefont {Lai}\ \emph {et~al.}(1994)\citenamefont {Lai},
  \citenamefont {Lai},\ and\ \citenamefont
  {Swinger}}]{lai_single-grating_1994}%
  \BibitemOpen
  \bibfield  {author} {\bibinfo {author} {\bibfnamefont {M.}~\bibnamefont
  {Lai}}, \bibinfo {author} {\bibfnamefont {S.~T.}\ \bibnamefont {Lai}},\ and\
  \bibinfo {author} {\bibfnamefont {C.}~\bibnamefont {Swinger}},\ }\bibfield
  {title} {\bibinfo {title} {Single-grating laser pulse stretcher and
  compressor},\ }\href {https://doi.org/10.1364/AO.33.006985} {\bibfield
  {journal} {\bibinfo  {journal} {Appl. Opt.}\ }\textbf {\bibinfo {volume}
  {33}},\ \bibinfo {pages} {6985} (\bibinfo {year} {1994})}\BibitemShut
  {NoStop}%
\bibitem [{\citenamefont {Martinez}(1986)}]{martinez1986grating}%
  \BibitemOpen
  \bibfield  {author} {\bibinfo {author} {\bibfnamefont {O.~E.}\ \bibnamefont
  {Martinez}},\ }\bibfield  {title} {\bibinfo {title} {Grating and prism
  compressors in the case of finite beam size},\ }\href
  {https://doi.org/https://doi.org/10.1364/JOSAB.3.000929} {\bibfield
  {journal} {\bibinfo  {journal} {J. Opt. Soc. Am. B}\ }\textbf {\bibinfo
  {volume} {3}},\ \bibinfo {pages} {929} (\bibinfo {year} {1986})}\BibitemShut
  {NoStop}%
\bibitem [{\citenamefont {Mukherjee}\ \emph {et~al.}(2020)\citenamefont
  {Mukherjee}, \citenamefont {Widhalm}, \citenamefont {Siebert}, \citenamefont
  {Krehs}, \citenamefont {Sharma}, \citenamefont {Thiede}, \citenamefont
  {Reuter}, \citenamefont {Förstner},\ and\ \citenamefont
  {Zrenner}}]{mukherjee_electrically_2020}%
  \BibitemOpen
  \bibfield  {author} {\bibinfo {author} {\bibfnamefont {A.}~\bibnamefont
  {Mukherjee}}, \bibinfo {author} {\bibfnamefont {A.}~\bibnamefont {Widhalm}},
  \bibinfo {author} {\bibfnamefont {D.}~\bibnamefont {Siebert}}, \bibinfo
  {author} {\bibfnamefont {S.}~\bibnamefont {Krehs}}, \bibinfo {author}
  {\bibfnamefont {N.}~\bibnamefont {Sharma}}, \bibinfo {author} {\bibfnamefont
  {A.}~\bibnamefont {Thiede}}, \bibinfo {author} {\bibfnamefont
  {D.}~\bibnamefont {Reuter}}, \bibinfo {author} {\bibfnamefont
  {J.}~\bibnamefont {Förstner}},\ and\ \bibinfo {author} {\bibfnamefont
  {A.}~\bibnamefont {Zrenner}},\ }\bibfield  {title} {\bibinfo {title}
  {Electrically controlled rapid adiabatic passage in a single quantum dot},\
  }\href {https://doi.org/10.1063/5.0012257} {\bibfield  {journal} {\bibinfo
  {journal} {Appl. Phys. Lett.}\ }\textbf {\bibinfo {volume} {116}},\ \bibinfo
  {pages} {251103} (\bibinfo {year} {2020})}\BibitemShut {NoStop}%
\bibitem [{\citenamefont {Hui}\ and\ \citenamefont
  {Liu}(2008)}]{hui_proposal_2008}%
  \BibitemOpen
  \bibfield  {author} {\bibinfo {author} {\bibfnamefont {H.~Y.}\ \bibnamefont
  {Hui}}\ and\ \bibinfo {author} {\bibfnamefont {R.~B.}\ \bibnamefont {Liu}},\
  }\bibfield  {title} {\bibinfo {title} {Proposal for geometric generation of a
  biexciton in a quantum dot using a chirped pulse},\ }\href
  {https://doi.org/10.1103/PhysRevB.78.155315} {\bibfield  {journal} {\bibinfo
  {journal} {Phys. Rev. B}\ }\textbf {\bibinfo {volume} {78}},\ \bibinfo
  {pages} {155315} (\bibinfo {year} {2008})}\BibitemShut {NoStop}%
\bibitem [{\citenamefont {Malinovsky}\ and\ \citenamefont
  {Krause}(2001)}]{malinovsky_general_2001}%
  \BibitemOpen
  \bibfield  {author} {\bibinfo {author} {\bibfnamefont {V.}~\bibnamefont
  {Malinovsky}}\ and\ \bibinfo {author} {\bibfnamefont {J.}~\bibnamefont
  {Krause}},\ }\bibfield  {title} {\bibinfo {title} {General theory of
  population transfer by adiabatic rapid passage with intense, chirped laser
  pulses},\ }\href {https://doi.org/10.1007/s100530170212} {\bibfield
  {journal} {\bibinfo  {journal} {Eur. Phys. J. D}\ }\textbf {\bibinfo {volume}
  {14}},\ \bibinfo {pages} {147} (\bibinfo {year} {2001})}\BibitemShut
  {NoStop}%
\bibitem [{\citenamefont {Glässl}\ \emph
  {et~al.}(2013{\natexlab{b}})\citenamefont {Glässl}, \citenamefont {Barth},
  \citenamefont {Gawarecki}, \citenamefont {Machnikowski}, \citenamefont
  {Croitoru}, \citenamefont {Lüker}, \citenamefont {Reiter}, \citenamefont
  {Kuhn},\ and\ \citenamefont {Axt}}]{glassl_biexciton_2013}%
  \BibitemOpen
  \bibfield  {author} {\bibinfo {author} {\bibfnamefont {M.}~\bibnamefont
  {Glässl}}, \bibinfo {author} {\bibfnamefont {A.~M.}\ \bibnamefont {Barth}},
  \bibinfo {author} {\bibfnamefont {K.}~\bibnamefont {Gawarecki}}, \bibinfo
  {author} {\bibfnamefont {P.}~\bibnamefont {Machnikowski}}, \bibinfo {author}
  {\bibfnamefont {M.~D.}\ \bibnamefont {Croitoru}}, \bibinfo {author}
  {\bibfnamefont {S.}~\bibnamefont {Lüker}}, \bibinfo {author} {\bibfnamefont
  {D.~E.}\ \bibnamefont {Reiter}}, \bibinfo {author} {\bibfnamefont
  {T.}~\bibnamefont {Kuhn}},\ and\ \bibinfo {author} {\bibfnamefont {V.~M.}\
  \bibnamefont {Axt}},\ }\bibfield  {title} {\bibinfo {title} {Biexciton state
  preparation in a quantum dot via adiabatic rapid passage: {Comparison}
  between two control protocols and impact of phonon-induced dephasing},\
  }\href {https://doi.org/10.1103/PhysRevB.87.085303} {\bibfield  {journal}
  {\bibinfo  {journal} {Phys. Rev. B}\ }\textbf {\bibinfo {volume} {87}},\
  \bibinfo {pages} {085303} (\bibinfo {year} {2013}{\natexlab{b}})}\BibitemShut
  {NoStop}%
\bibitem [{\citenamefont {Wei}\ \emph {et~al.}(2014)\citenamefont {Wei},
  \citenamefont {He}, \citenamefont {Chen}, \citenamefont {Hu}, \citenamefont
  {He}, \citenamefont {Wu}, \citenamefont {Schneider}, \citenamefont {Kamp},
  \citenamefont {Höfling}, \citenamefont {Lu},\ and\ \citenamefont
  {Pan}}]{wei_deterministic_2014}%
  \BibitemOpen
  \bibfield  {author} {\bibinfo {author} {\bibfnamefont {Y.-J.}\ \bibnamefont
  {Wei}}, \bibinfo {author} {\bibfnamefont {Y.-M.}\ \bibnamefont {He}},
  \bibinfo {author} {\bibfnamefont {M.-C.}\ \bibnamefont {Chen}}, \bibinfo
  {author} {\bibfnamefont {Y.-N.}\ \bibnamefont {Hu}}, \bibinfo {author}
  {\bibfnamefont {Y.}~\bibnamefont {He}}, \bibinfo {author} {\bibfnamefont
  {D.}~\bibnamefont {Wu}}, \bibinfo {author} {\bibfnamefont {C.}~\bibnamefont
  {Schneider}}, \bibinfo {author} {\bibfnamefont {M.}~\bibnamefont {Kamp}},
  \bibinfo {author} {\bibfnamefont {S.}~\bibnamefont {Höfling}}, \bibinfo
  {author} {\bibfnamefont {C.-Y.}\ \bibnamefont {Lu}},\ and\ \bibinfo {author}
  {\bibfnamefont {J.-W.}\ \bibnamefont {Pan}},\ }\bibfield  {title} {\bibinfo
  {title} {Deterministic and {Robust} {Generation} of {Single} {Photons} from a
  {Single} {Quantum} {Dot} with 99.5\% {Indistinguishability} {Using}
  {Adiabatic} {Rapid} {Passage}},\ }\href {https://doi.org/10.1021/nl503081n}
  {\bibfield  {journal} {\bibinfo  {journal} {Nano Lett.}\ }\textbf {\bibinfo
  {volume} {14}},\ \bibinfo {pages} {6515} (\bibinfo {year}
  {2014})}\BibitemShut {NoStop}%
\bibitem [{\citenamefont {Lüker}\ \emph {et~al.}(2012)\citenamefont {Lüker},
  \citenamefont {Gawarecki}, \citenamefont {Reiter}, \citenamefont
  {Grodecka-Grad}, \citenamefont {Axt}, \citenamefont {Machnikowski},\ and\
  \citenamefont {Kuhn}}]{luker_influence_2012}%
  \BibitemOpen
  \bibfield  {author} {\bibinfo {author} {\bibfnamefont {S.}~\bibnamefont
  {Lüker}}, \bibinfo {author} {\bibfnamefont {K.}~\bibnamefont {Gawarecki}},
  \bibinfo {author} {\bibfnamefont {D.~E.}\ \bibnamefont {Reiter}}, \bibinfo
  {author} {\bibfnamefont {A.}~\bibnamefont {Grodecka-Grad}}, \bibinfo {author}
  {\bibfnamefont {V.~M.}\ \bibnamefont {Axt}}, \bibinfo {author} {\bibfnamefont
  {P.}~\bibnamefont {Machnikowski}},\ and\ \bibinfo {author} {\bibfnamefont
  {T.}~\bibnamefont {Kuhn}},\ }\bibfield  {title} {\bibinfo {title} {Influence
  of acoustic phonons on the optical control of quantum dots driven by
  adiabatic rapid passage},\ }\href
  {https://doi.org/10.1103/PhysRevB.85.121302} {\bibfield  {journal} {\bibinfo
  {journal} {Phys. Rev. B}\ }\textbf {\bibinfo {volume} {85}},\ \bibinfo
  {pages} {121302} (\bibinfo {year} {2012})}\BibitemShut {NoStop}%
\bibitem [{\citenamefont {Lüker}\ \emph {et~al.}(2017)\citenamefont {Lüker},
  \citenamefont {Kuhn},\ and\ \citenamefont {Reiter}}]{luker_phonon_2017}%
  \BibitemOpen
  \bibfield  {author} {\bibinfo {author} {\bibfnamefont {S.}~\bibnamefont
  {Lüker}}, \bibinfo {author} {\bibfnamefont {T.}~\bibnamefont {Kuhn}},\ and\
  \bibinfo {author} {\bibfnamefont {D.~E.}\ \bibnamefont {Reiter}},\ }\bibfield
   {title} {\bibinfo {title} {Phonon impact on optical control schemes of
  quantum dots: {Role} of quantum dot geometry and symmetry},\ }\href
  {https://doi.org/10.1103/PhysRevB.96.245306} {\bibfield  {journal} {\bibinfo
  {journal} {Phys. Rev. B}\ }\textbf {\bibinfo {volume} {96}},\ \bibinfo
  {pages} {245306} (\bibinfo {year} {2017})}\BibitemShut {NoStop}%
\bibitem [{\citenamefont {Reiter}\ \emph {et~al.}(2019)\citenamefont {Reiter},
  \citenamefont {Kuhn},\ and\ \citenamefont {Axt}}]{reiter_distinctive_2019}%
  \BibitemOpen
  \bibfield  {author} {\bibinfo {author} {\bibfnamefont {D.~E.}\ \bibnamefont
  {Reiter}}, \bibinfo {author} {\bibfnamefont {T.}~\bibnamefont {Kuhn}},\ and\
  \bibinfo {author} {\bibfnamefont {V.~M.}\ \bibnamefont {Axt}},\ }\bibfield
  {title} {\bibinfo {title} {Distinctive characteristics of carrier-phonon
  interactions in optically driven semiconductor quantum dots},\ }\href
  {https://doi.org/10.1080/23746149.2019.1655478} {\bibfield  {journal}
  {\bibinfo  {journal} {Adv. Phys.}\ }\textbf {\bibinfo {volume} {4}},\
  \bibinfo {pages} {1655478} (\bibinfo {year} {2019})}\BibitemShut {NoStop}%
\bibitem [{\citenamefont {Huber}\ \emph {et~al.}(2017)\citenamefont {Huber},
  \citenamefont {Reindl}, \citenamefont {Huo}, \citenamefont {Huang},
  \citenamefont {Wildmann}, \citenamefont {Schmidt}, \citenamefont {Rastelli},\
  and\ \citenamefont {Trotta}}]{huber2017highly}%
  \BibitemOpen
  \bibfield  {author} {\bibinfo {author} {\bibfnamefont {D.}~\bibnamefont
  {Huber}}, \bibinfo {author} {\bibfnamefont {M.}~\bibnamefont {Reindl}},
  \bibinfo {author} {\bibfnamefont {Y.}~\bibnamefont {Huo}}, \bibinfo {author}
  {\bibfnamefont {H.}~\bibnamefont {Huang}}, \bibinfo {author} {\bibfnamefont
  {J.~S.}\ \bibnamefont {Wildmann}}, \bibinfo {author} {\bibfnamefont {O.~G.}\
  \bibnamefont {Schmidt}}, \bibinfo {author} {\bibfnamefont {A.}~\bibnamefont
  {Rastelli}},\ and\ \bibinfo {author} {\bibfnamefont {R.}~\bibnamefont
  {Trotta}},\ }\bibfield  {title} {\bibinfo {title} {{Highly indistinguishable
  and strongly entangled photons from symmetric GaAs quantum dots}},\ }\href
  {https://doi.org/10.1038/ncomms15506} {\bibfield  {journal} {\bibinfo
  {journal} {Nat. Commun.}\ }\textbf {\bibinfo {volume} {8}},\ \bibinfo {pages}
  {15506} (\bibinfo {year} {2017})}\BibitemShut {NoStop}%
\bibitem [{\citenamefont {da~Silva}\ \emph {et~al.}(2021)\citenamefont
  {da~Silva}, \citenamefont {Undeutsch}, \citenamefont {Lehner}, \citenamefont
  {Manna}, \citenamefont {Krieger}, \citenamefont {Reindl}, \citenamefont
  {Schimpf}, \citenamefont {Trotta},\ and\ \citenamefont
  {Rastelli}}]{da2021gaas}%
  \BibitemOpen
  \bibfield  {author} {\bibinfo {author} {\bibfnamefont {S.~F.~C.}\
  \bibnamefont {da~Silva}}, \bibinfo {author} {\bibfnamefont {G.}~\bibnamefont
  {Undeutsch}}, \bibinfo {author} {\bibfnamefont {B.}~\bibnamefont {Lehner}},
  \bibinfo {author} {\bibfnamefont {S.}~\bibnamefont {Manna}}, \bibinfo
  {author} {\bibfnamefont {T.~M.}\ \bibnamefont {Krieger}}, \bibinfo {author}
  {\bibfnamefont {M.}~\bibnamefont {Reindl}}, \bibinfo {author} {\bibfnamefont
  {C.}~\bibnamefont {Schimpf}}, \bibinfo {author} {\bibfnamefont
  {R.}~\bibnamefont {Trotta}},\ and\ \bibinfo {author} {\bibfnamefont
  {A.}~\bibnamefont {Rastelli}},\ }\bibfield  {title} {\bibinfo {title} {{GaAs
  quantum dots grown by droplet etching epitaxy as quantum light sources}},\
  }\href {https://doi.org/10.1063/5.0057070} {\bibfield  {journal} {\bibinfo
  {journal} {Appl. Phys. Lett.}\ }\textbf {\bibinfo {volume} {119}},\ \bibinfo
  {pages} {120502} (\bibinfo {year} {2021})}\BibitemShut {NoStop}%
\bibitem [{\citenamefont {Duquennoy}\ \emph {et~al.}(2022)\citenamefont
  {Duquennoy}, \citenamefont {Colautti}, \citenamefont {Emadi}, \citenamefont
  {Majumder}, \citenamefont {Lombardi},\ and\ \citenamefont
  {Toninelli}}]{duquennoy_real-time_2022}%
  \BibitemOpen
  \bibfield  {author} {\bibinfo {author} {\bibfnamefont {R.}~\bibnamefont
  {Duquennoy}}, \bibinfo {author} {\bibfnamefont {M.}~\bibnamefont {Colautti}},
  \bibinfo {author} {\bibfnamefont {R.}~\bibnamefont {Emadi}}, \bibinfo
  {author} {\bibfnamefont {P.}~\bibnamefont {Majumder}}, \bibinfo {author}
  {\bibfnamefont {P.}~\bibnamefont {Lombardi}},\ and\ \bibinfo {author}
  {\bibfnamefont {C.}~\bibnamefont {Toninelli}},\ }\bibfield  {title} {\bibinfo
  {title} {Real-time two-photon interference from distinct molecules on the
  same chip},\ }\href {https://doi.org/10.1364/OPTICA.452317} {\bibfield
  {journal} {\bibinfo  {journal} {Optica}\ }\textbf {\bibinfo {volume} {9}},\
  \bibinfo {pages} {731} (\bibinfo {year} {2022})}\BibitemShut {NoStop}%
\bibitem [{\citenamefont {Koong}\ \emph {et~al.}(2020)\citenamefont {Koong},
  \citenamefont {Ballesteros-Garcia}, \citenamefont {Proux}, \citenamefont
  {Dalacu}, \citenamefont {Poole},\ and\ \citenamefont
  {Gerardot}}]{koong_multiplexed_2020}%
  \BibitemOpen
  \bibfield  {author} {\bibinfo {author} {\bibfnamefont {Z.-X.}\ \bibnamefont
  {Koong}}, \bibinfo {author} {\bibfnamefont {G.}~\bibnamefont
  {Ballesteros-Garcia}}, \bibinfo {author} {\bibfnamefont {R.}~\bibnamefont
  {Proux}}, \bibinfo {author} {\bibfnamefont {D.}~\bibnamefont {Dalacu}},
  \bibinfo {author} {\bibfnamefont {P.~J.}\ \bibnamefont {Poole}},\ and\
  \bibinfo {author} {\bibfnamefont {B.~D.}\ \bibnamefont {Gerardot}},\
  }\bibfield  {title} {\bibinfo {title} {Multiplexed {Single} {Photons} from
  {Deterministically} {Positioned} {Nanowire} {Quantum} {Dots}},\ }\href
  {https://doi.org/10.1103/PhysRevApplied.14.034011} {\bibfield  {journal}
  {\bibinfo  {journal} {Phys. Rev. Appl.}\ }\textbf {\bibinfo {volume} {14}},\
  \bibinfo {pages} {034011} (\bibinfo {year} {2020})}\BibitemShut {NoStop}%
\bibitem [{\citenamefont {Glässl}\ \emph {et~al.}(2011)\citenamefont
  {Glässl}, \citenamefont {Vagov}, \citenamefont {Lüker}, \citenamefont
  {Reiter}, \citenamefont {Croitoru}, \citenamefont {Machnikowski},
  \citenamefont {Axt},\ and\ \citenamefont {Kuhn}}]{glassl_long-time_2011}%
  \BibitemOpen
  \bibfield  {author} {\bibinfo {author} {\bibfnamefont {M.}~\bibnamefont
  {Glässl}}, \bibinfo {author} {\bibfnamefont {A.}~\bibnamefont {Vagov}},
  \bibinfo {author} {\bibfnamefont {S.}~\bibnamefont {Lüker}}, \bibinfo
  {author} {\bibfnamefont {D.~E.}\ \bibnamefont {Reiter}}, \bibinfo {author}
  {\bibfnamefont {M.~D.}\ \bibnamefont {Croitoru}}, \bibinfo {author}
  {\bibfnamefont {P.}~\bibnamefont {Machnikowski}}, \bibinfo {author}
  {\bibfnamefont {V.~M.}\ \bibnamefont {Axt}},\ and\ \bibinfo {author}
  {\bibfnamefont {T.}~\bibnamefont {Kuhn}},\ }\bibfield  {title} {\bibinfo
  {title} {Long-time dynamics and stationary nonequilibrium of an optically
  driven strongly confined quantum dot coupled to phonons},\ }\href
  {https://doi.org/10.1103/PhysRevB.84.195311} {\bibfield  {journal} {\bibinfo
  {journal} {Phys. Rev. B}\ }\textbf {\bibinfo {volume} {84}},\ \bibinfo
  {pages} {195311} (\bibinfo {year} {2011})}\BibitemShut {NoStop}%
\end{thebibliography}%

\end{document}